\documentclass[review]{elsarticle}

\usepackage{lineno,hyperref}
\modulolinenumbers[5]

\journal{Journal of \LaTeX\ Templates}
\usepackage[font=small]{caption}
\usepackage{ifpdf}
\usepackage{commath}
\usepackage{multicol}
\usepackage{lipsum} 
\usepackage{cuted} 
\usepackage{stix}
\usepackage{pgfplots}
\usetikzlibrary{calc,arrows,decorations.markings}
\usepackage{bigints}
\newif\ifCLASSOPTIONonecolumn       \CLASSOPTIONonecolumnfalse
\newif\ifCLASSOPTIONtwocolumn       \CLASSOPTIONtwocolumntrue
\usepackage{caption}
\usepackage{subcaption}
\usepackage{amsmath}
\usepackage{dsfont}
\usepackage{bbold}
\usepackage{amssymb}
\usepackage{algorithmic}
\usepackage{array}
\usepackage{stfloats}
\usepackage{bigints}
\journal{Arxiv}

\begin{document}

\begin{frontmatter}

\title{On the Performance of Cooperative NOMA Using MRC at Road Intersections in the Presence of Interference\tnoteref{fund}}

\author[n7]{Baha Eddine Youcef~Belmekki\corref{cor1}}
\ead{bahaeddine.belmekki@enseeiht.fr}
\author[usthb]{Abdelkrim ~Hamza}
\ead{ahamza@usthb.dz}
\author[n7]{Beno\^it~Escrig}
\ead{benoit.escrig@enseeiht.fr}
\address[n7]{University of Toulouse, IRIT/INP-ENSEEIHT Toulouse, BP 7122, 31071 Toulouse Cedex 7, France}
\address[usthb]{LISIC Laboratory, Electronic and Computer Faculty, USTHB, Algiers, Algeria}
\cortext[cor1]{Corresponding author}

\tnotetext[fund]{Results related to this paper have been presented at the wireless and mobile computing, networking and communications (WiMob) 2019, Barcelona, Spain, October 2019 \cite{Belmkki}.}

\begin{abstract}
As the traffic safety has become of utmost importance, much attention is given to
intelligent transportation systems (ITSs), and more particularly to vehicular communications (VCs). Moreover, 50 \% of all crashes
happen at road intersections, which makes theme a critical areas.
In this paper, we investigate the improvement when implementing maximum ratio combining (MRC) in cooperative VCs transmission schemes using non-orthogonal multiple access scheme (NOMA) at road intersections. 
We consider that a source transmits a message to two destinations with a aid of a relay.
The transmission undergoes interference generated from a set of vehicles on the roads. 
We obtained closed form outage probability expressions, and we extend the derivation for a scenario involving $K$ destination nodes and several road lanes. 
The performance of MRC cooperative NOMA is compared with the standard cooperative NOMA, and we show that implementing MRC with NOMA offers a significant improvement over the standard cooperative NOMA. Also, we compare the performance of MRC using NOMA with MRC cooperative orthogonal multiple access (OMA), and demonstrate that NOMA significantly outperforms OMA. We conclude that it is always beneficial to use MRC and NOMA even at the cost of implementation complexity. 
Finally, we demonstrate that the outage probability increases drasticallyen the vehicles are closer to the road intersection, and that using MRC with NOMA improves significantly the performance in this context. 
To verify the correctness of our analysis, extensive Monte-Carlo simulations are carried out.

\end{abstract}

\begin{keyword}
5G, NOMA, interference, outage probability, cooperative, MRC, intersections.
\end{keyword}

\end{frontmatter}

\linenumbers

\section{Introduction}
\subsection{Motivation}
As the traffic safety has become of utmost importance, much attention is given to intelligent transportation systems (ITSs), and more specifically to vehicular communications (VCs) \cite{arif2019survey,mekki2017vehicular}. VCs offer a wide range of applications such as, traffic state, autonomous driving, and safety \cite{sam2016vehicle,singh2019multipath,boquet2018adaptive}. According to World Health organization, over 1.25 million people die each year on the roads, and road traffic crashes are the number one cause of death among young people \cite{world2015global}. Moreover, 50 \% of all crashes are in junction areas (intersections) including fatal crashes, injury crashes and property damage crashes \cite{traficsafety}. This makes intersections critical areas not only for vehicles, but also for pedestrians and cyclists.
VCs have numerous applications to prevent accidents, or alert vehicles when accidents happen in their surroundings. 
Thus, low latency and high reliability communications are mandatory in safety-based VCs. 
To increase spectral efficiency and data rate \cite{ding2017application} in the fifth generation (5G) of wireless communication systems, non-orthogonal multiple access (NOMA) is an promising candidate as a multiple access scheme. 
NOMA, unlike orthogonal multiple access (OMA), allows several users to use the same resource with several power allocation levels.
Also, cooperative transmissions have been show to increase the reliability of the transmission link \cite{wang2018secure,nguyen2019performance,altieri2019performance}. On the other hand, co-channel interference is one of the major impairments that can degrade a transmission in VCs \cite{tripp2018comparison,mourad2017performance,campolo2016modeling}. 
Hence, in this paper we propose to study the impact of interference in cooperative VCs at intersections using NOMA.

\subsection{Related Works}
\subsubsection{NOMA Works}

The performance of NOMA has been well studied in the literature (see \cite{dai2015non,islam2017power,ding2014performance} and the references therein). 
As far as the impact of interference on NOMA is concerned, several papers have studied its effect \cite{zhang2016stochastic}. The authors in \cite{zhang2017uplink} analysed the impact of interference on a NOMA uplink transmission. The authors also analyzed the performance of a NOMA downlink transmission with a selection based pairing in \cite{zhang2017downlink}.
The improvement of using cooperative transmissions  in NOMA have been also well investigated \cite{ding2016relay,liu2016cooperative,timotheou2015fairness,ding2014performance}. 
A scenario involving $M$ number of randomly deployed users was investigated in \cite{ding2014performance}. The authors also evaluated the ergodic rate and outage performance in \cite{timotheou2015fairness}.
In \cite{ding2016relay}, the authors studied the impact of relay selection on cooperative NOMA, and showed that the two-stage scheme can achieve the optimal diversity gain and the minimal outage probability.
However, the impact of implementing NOMA into VCs has been lacking in the literature.

\subsubsection{VCs Works}
The performance of VCs in the presence of interference has attracted a lot of attention \cite{rakhshan2017improving,jiang2016information,farooq2016stochastic}. 
Mainly, there are two types of scenarios in VCs, highways scenarios and intersections scenarios. Considering highway scenarios, the authors in \cite{jiang2016information} investigated the performance of RTS/CTS protocol considering Nakagami-$m$ channels fading. In \cite{rakhshan2017improving}, the authors  studied how the interference affects the safety of vehicles in a VCs. The authors also derived the packet success probability for two different traffic models in VCs \cite{rakhshan2016packet}.
The authors in \cite{farooq2016stochastic} investigated the performance of carrier sense multiple access (CSMA) protocols, and derived the expressions of packet success probability. In \cite{tassi2017modeling}, the authors derived the outage probability and rate coverage probability when a line of sight path to the base station is absent.

Considering intersection scenarios, a success probability expression of a simple intersection scenario was derived in \cite{steinmetz2015stochastic}. The authors in \cite{abdulla2016vehicle} extended the work of \cite{steinmetz2015stochastic} and derived the success probability considering limited road segments with different path loss models. The authors of \cite{abdulla2016vehicle} also studied the average and the fine-grained reliability in an interference-limited vehicle to vehicle (V2V) communications with the aid of the meta distribution in \cite{abdulla2017fine}. The authors \cite{jeyaraj2017reliability} in investigated the performance of V2V communications for orthogonal streets. The authors also studied V2V communications at intersections and showed that, the performance of the ALOHA protocol can be considered as lower bound of performance of the CSMA protocols \cite{jeyaraj2018nearest}. The effect of vehicles mobility and interference dependence has been investigated in \cite{J2}. The authors also, studied the performance of three transmission schemes at intersection in line of sight scenario and non light of sight scenario considering Nakagami-$m$ fading channels in \cite{J1,belmekki2019outagearxiv}.

However, the performance of NOMA in VCs is lacking in the literature. The first to tackle this issue are the authors of the paper at hand. They computed the outage probability and average achievable rate of NOMA at intersection roads considering direct transmissions \cite{C1,J3} and cooperative transmissions \cite{belmekki2019outage,J4}. They also investigated the performance of NOMA in millimeter wave vehicular communications in \cite{Cmm1,Cmm2}. In \cite{bprotocol}, the authors proposed an adaptive NOMA protocol in VCs.

In this paper, the authors study the feasibility and improvement in performance by implementing both NOMA and maximum ratio combining (MRC) in VCs. Hence, we compare the proposed scheme with the classical OMA, and the classical cooperative NOMA, and see if the improvements justify and outweigh the complexity of implementing MRC and NOMA in VCs.

\subsection{Contributions}
The  contributions of this paper are  as follows:
\begin{itemize}
\item We establish a framework for performance analysis of VCs under Rayleigh fading and two perpendicular roads containing one-dimensional Poisson field of interference. We analyze the performance of implementing  MRC in cooperative VCs transmission schemes using NOMA at intersections in terms of outage probability. We obtained closed form outage probability 
expressions. We further extend the derivations when $K$ destination nodes are involved, and to a realistic intersection scenario involving multiple lanes.

\item We compare the performance of MRC cooperative NOMA with a classical cooperative NOMA \cite{J4}, and show that implementing MRC in cooperative NOMA transmission offers a significant improvement over the classical cooperative NOMA in terms of outage probability. We also compare the performance of MRC cooperative NOMA with MRC cooperative OMA \cite{J2}, and show that NOMA offers a better performance than OMA. It is shown that the outage probability increases when the vehicles are closer to the road intersection, and that using MRC considering NOMA improves significantly the performance in this context. 

\item
The relationships between system performance
and different network parameters such as NOMA power allocation coefficient, date rates,
channel access probability, intensity of potential interfering vehicles, relay position, noise power levels, successive interference cancellation (SIC) coefficient
are discussed. The
results clearly demonstrate the advantages of implementing MRC into NOMA
the performance in VCs, even at the cost of implementation complexity.

\item We show that as we increases the data rate of $D_2$, MRC transmission using NOMA offers a better performance than MRC transmission using OMA. Whereas for $D_1 $, low data rates are suitable, since there is a condition imposed to its data rate. We also show how the imperfect SIC process can degrade the performance of NOMA. We also show that MRC transmission using NOMA outperforms cooperative NOMA. 
Finally, we investigate the best relay position, and show that the optimal relay position for $D_1$ and $D_2$ is near the destination nodes.

\item To confirm the correctness of our theoretical derivations, extensive Monte-Carlo simulations are carried out.

\end{itemize}

\subsection{Organization}

The rest of this paper is organized as follows. Section \ref{Section2} presents the system model. In Section \ref{Section3}, outage analytical expressions are derived. The Laplace transform expressions are presented in Section \ref{Section4}. Extension to multiple lanes scenario is investigated in Section \ref{Section5}. Simulations and discussions are in Section \ref{Section6}. Finally, we conclude the paper in Section \ref{Section7}.

\section{System Model}\label{Section2}
\begin{figure}[]
\centering
\includegraphics[scale=0.65]{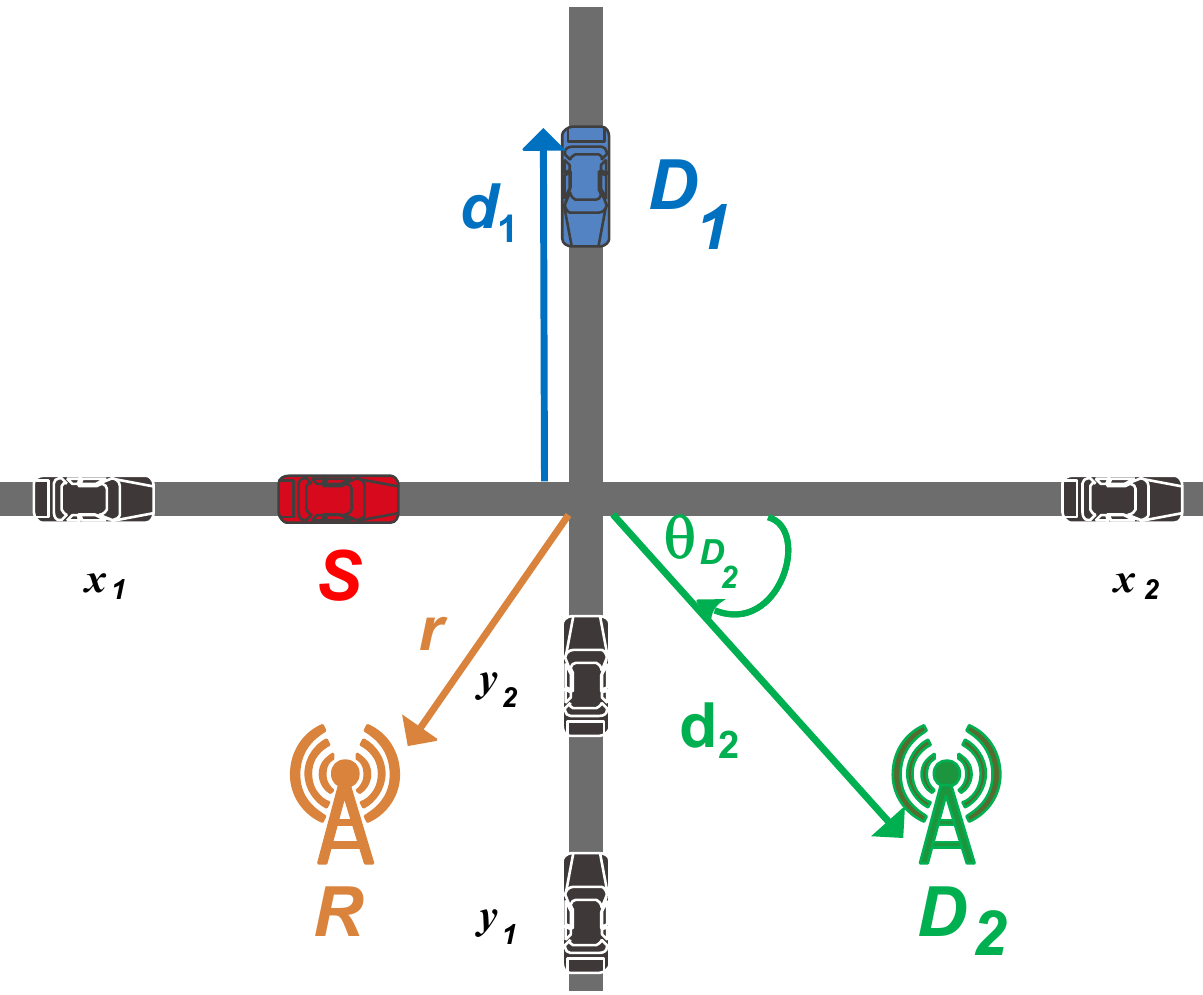}
\caption{Cooperative NOMA system model for vehicular communications involving two destination nodes and a relay node. For this example, $S$ is a vehicle, $R$ is an infrastructure, $D_1$ is a vehicle, and $D_2$ is an infrastructure. }
\label{Figure1}
\end{figure}

\subsection{Intersection Scenario}
We consider a cooperative transmission using NOMA between a source $S$ and two destinations $D_1$ and $D_2$, with the aid of a relay $R$ as shown in Fig.\ref{Figure1}. As both V2V and V2I communications are of interest\footnote{The Doppler shift and time-varying effect of V2V and V2I channels is beyond the scope of this paper.}, the nodes $S$,$R$, $D_1$ and $D_2$ can be on the roads (as vehicles), or outside the roads (as infrastructures). For instance in Fig.\ref{Figure1}, the configuration is as follows: $S$ and $D_1$ are vehicles, whereas $R$ and $D_2$ are infrastructures. For the sake of notation simplicity, we denote by $M$ the receiving node, and by $m$ the distance between the node $M$ and the intersection, where $M \in \{R,D_1,D_2\}$ and $m \in \{r,d_1,d_2\}$, as shown in Fig.\ref{Figure1}. Also, the term $\theta_M $ denotes the angle between the node $M$ and the $X$ road.

In this paper, we study the performance at an intersection. The intersection has two two perpendicular roads, an horizontal road denoted by $X$, and a vertical road denoted by $Y$. We extend the analysis to the case when the intersection involves multiple lanes in Section \ref{Section5}.

The set of nodes $\lbrace{S, R, D_1, D_2}\rbrace$ is subject to interference originated from transmitting vehicles located on the roads. 
The set of interfering vehicles located on the $Z$ road where $Z\in \{X,Y\}$, denoted by $\Phi_{Z}$ are modeled as a one-dimensional homogeneous Poisson point process (1D-HPPP), that is, $\Phi_{Z}\sim\textrm{1D-HPPP}(\lambda_{Z},z)$, where $z\in \{x,y\}$ and $\lambda_{Z}$ are the position of interfering vehicles and their intensity on the $Z$ road, respectively. This implies that the number of potential interfering vehicles within any closed and bounded set $\mathcal{B}\subseteq\mathbb{R}$ is a Poisson random variable with parameter $\lambda \vert \mathcal{B}\vert $.

\begin{figure}
\centering
\fbox{%
\fbox{%

\begin{tikzpicture}

\tikzstyle{S}=[circle,very thick,draw=black, fill=red, minimum size=.7cm]
\tikzstyle{R}=[circle,very thick,draw=black, fill=orange, minimum size=.7cm]
\tikzstyle{D1}=[circle,very thick,draw=black, fill=blue, minimum size=.7cm]
\tikzstyle{D2}=[circle,very thick,draw=black, fill=green, minimum size=.7cm]

\node[S,label=below :{$S$}] (S) at (0,0) {};
\node[R,label=below :{$R$}] (R) at (2, 0) {};
\node[D1,label=below :{$D_1$}] (D1) at (4, 2) {};
\node[D2,label=below :{$D_2$}] (D2) at (4, -2) {};

\draw [->, very thick] (S) -- (R); \draw [->, very thick](S) -- (D1) ;\draw [->, very thick](S) -- (D2) ;

\node[draw,text width=3cm] at (2,4) {\textbf{First phase}};
\end{tikzpicture}
}
\hspace{1cm}
\fbox{%

\begin{tikzpicture}

\tikzstyle{S}=[circle,very thick,draw=black, fill=red, minimum size=.7cm]
\tikzstyle{R}=[circle,very thick,draw=black, fill=orange, minimum size=.7cm]
\tikzstyle{D1}=[circle,very thick,draw=black, fill=blue, minimum size=.7cm]
\tikzstyle{D2}=[circle,very thick,draw=black, fill=green, minimum size=.7cm]

\node[S,label=below :{$S$}] (S) at (6,0) {};
\node[R,label=below :{$R$}] (R) at (8, 0) {};
\node[D1,label=below :{$D_1$}] (D1) at (10, 2) {};
\node[D2,label=below :{$D_2$}] (D2) at (10, -2) {};

\draw [->, very thick](R) -- (D1); \draw [->, very thick](R) -- (D2);

\node[draw,text width=3cm] at (8,4) {\textbf{Second phase}};
\end{tikzpicture}
}
}
\caption{Transmission scheme using MRC and NOMA.} \label{fig:M1}
\end{figure}
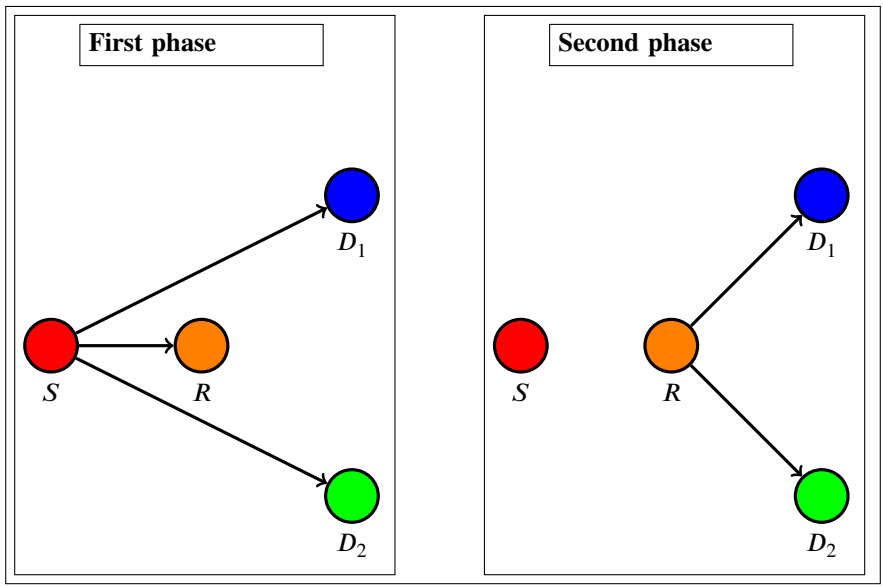

\subsection{MRC and Cooperative Protocol}
In this paper, we use Decode and Forward (DF) cooperative protocol \cite{feteiha2016decode,altieri2014outage}. The transmission occurs in two phases, the duration of each phase is one time slot. Finally we consider we use MRC in NOMA setup as shown in Fig.\ref{fig:M1}. In the first phase, $S$ broadcasts the message, and the nodes $R$, $D_1$ and $D_2$ try to decode the message. In the second phase, if $R$ decodes $S$ message, it broadcasts the message to $D_1$ and $D_2$. Then, $D_1$ and $D_2$ add the power received in the first phase from $S$ and (if $R$ decodes $S$ message) the power received from $R$ during the second phase to decode the message.

\subsection{NOMA Scenario and Assumptions}
In NOMA, there are two main ways to order the users. The first one is to order the nodes according to their channel stats. Hence, the user with the weakest channel state comes first in the decoding order (see \cite{ding2014performance,ding2015cooperative} and references therein). The second one is that, the users are sorted according to their quality of service (QoS) priorities. Hence, a user with the higher priority comes first in the seconding order. It has been show in \cite{ding2016relay,ding2016mimo}, that ordering users according to their QoS is more realistic and reasonable assumption, since in practice, it  is  very  likely  that users who want to participate in NOMA have similar channel conditions. Without loss of generality, we study the case in which node $D_1$ has to be served immediately with a low data rate.  For example, $D_1$ can be a vehicle that needs to receive safety information containing  a  few  bytes, such as a road flood warning or incident avoidance alert message. Whereas node $D_2$ requires relatively high data rate but can be served later. For instance $D_2$ can be a user that accesses the internet connection.

\subsection{Channel Model}
We consider slotted ALOHA protocol with parameter $p$, i.e., every node can access the medium with a probability $p$. This performs an independent thinning the parent 1D-HPPP by probability $p$. Hence, the set of interfering vehicles at a given time slot also follow a a 1D-HPPP with intensity $p\lambda$.

The transmission between a node $a$ and $b$ experience a path loss given by $l_{ab}= (A r_{ab})^{-\alpha}$
, where $A$ is a constant depending on the antenna characteristics, $ r_{ab}=\Vert a-b\Vert$, and $\alpha$ is the path loss exponent. All the node transmit with power $P$.

The signal transmitted by $S$, denoted $ \chi_{S}$ is a mixture of the message intended to $D_1$ and $D_2$. This can be expressed as
\begin{equation}
 \chi_{S}=\sqrt{a_1}\chi_{D1}+\sqrt{a_2}\chi_{D2}, \nonumber
 \end{equation}
where $a_i$ is the power coefficients allocated to $D_i$, and $\chi_{Di}$ is the message intended to $D_i$, where $i \in \{1,2\}$. Since $D_1$ has higher power than $D_2$, that is $a_1 \ge a_2$, then $D_1$ comes first in the decoding order. Note that, $a_1+a_2=1$.\\
 The signal received at $R$ and $D_i$, denoted respectively by $\mathcal{Y}_{R}$ and $\mathcal{Y}_{D_i}$, during the first time slot are expressed as
 \begin{flalign}
 &\mathcal{Y}_{R}=\underbrace{h_{SR}\sqrt{P l_{SR}}\:\chi_{S}}_\text{The signal of interest that contains $D_1$ message and $D_2$ message}+\nonumber \\
 &\underbrace{\sum_{x\in \Phi_{X_{R}}}h_{Rx}\sqrt{P l_{Rx}}\:\chi_{x}}_\text{Aggregate interference form the $Y$ road at $R$} 
 +\underbrace{\sum_{y\in \Phi_{Y_{R}}}h_{Ry}\sqrt{P l_{Ry}}\:\chi_{y}}_\text{Aggregate interference form the $Y$ road at $R$}+\underbrace{\sigma^2}_\text{Noise related term}, \nonumber
 \end{flalign}
 
 and

 \begin{flalign}
  & \mathcal{Y}_{D_i}=h_{SD_i}\sqrt{P l_{SD_i}}\:\chi_{S}+ \nonumber \\
 &\underbrace{\sum_{x\in \Phi_{X_{D_i}}}h_{D_ix}\sqrt{P l_{D_ix}}\:\chi_{x}}_\text{Aggregate interference form the $X$ road at $D_i$}
 + \underbrace{\sum_{y\in \Phi_{Y_{D_i}}}h_{D_iy}\sqrt{P l_{D_iy}}\:\chi_{y}}_\text{Aggregate interference form the $Y$ road at $D_i$}+\sigma^2. \nonumber
 \end{flalign}
 The signal received at $D_i$, denoted by $\mathcal{Y}_{D_i}$, during the second time slot is expressed as
 \begin{flalign}
   &\mathcal{Y}_{D_i}=h_{RD_i}\sqrt{P l_{RD_i}}\:\chi_{R}+ \nonumber \\
 &\underbrace{\sum_{x\in \Phi_{X_{D_i}}}h_{D_ix}\sqrt{P l_{D_ix}}\:\chi_{x}}_\text{Aggregate interference form the $Y$ road at $D_i$} 
 +\underbrace{\sum_{y\in \Phi_{Y_{D_i}}}h_{D_iy}\sqrt{P l_{D_iy}}\:\chi_{y}}_\text{Aggregate interference form the $Y$ road at $D_i$}+\sigma^2, \nonumber
 \end{flalign}
where signals transmitted by the interfering vehicles $x$ and $y$, are denoted by $ \chi_x$ and $\chi_y $, respectively. The term $h_{ab}$ denotes the fading coefficient between node $a$ and $b$, and it is modeled as $\mathcal{CN}(0,1)$ \cite{halimi2017wavelet1,halimi2017unsupervised2,halimi2017statistical3}, hence $|h_{ab}|^2 \sim \exp(1)$. The aggregate interference is defined as 
\begin{equation}\label{eqation.1}
I_{Z_{M}}=\sum_{z\in \Phi_{Z_{M}}}P\vert h_{Mz}\vert^{2}l_{Mz} , 
\end{equation}
where $I_{Z_{M}} $ denotes the aggregate interference from the $Z$ road at $M$, $\Phi_{Z_{M}}$ denotes the set of the interfering vehicles from the $Z$ road at $M$.

\section{Outage Analytical Derivations}\label{Section3}
\subsection{Outage Events}
We define an outage event at the receiving node when the signal-to-interference plus noise ratio (SINR) at the receiver is below a given threshold. According to SIC \cite{hasna2003performance}, $D_1$ is decoded first since it has the higher power allocation, and $D_2$ message is considered as interference. The outage event at $R$ to not decode $D_1$, denoted  $\mathcal{A}_{R_1}(\Theta_1)$, is defined as
\begin{equation}
\mathcal{A}_{R_1}(\Theta_1)\triangleq \frac{P\vert  h_{SR}\vert^{2}l_{SR}\,a_1}{P\vert h_{SR}\vert^{2}l_{SR}a_2+I_{X_{R}}+I_{Y_{R}}+\sigma^2} < \Theta_1,
\end{equation}
where $\Theta_1=2^{2\mathcal{R}_1}-1$, and $\mathcal{R}_1$ is the target data rate of $D_1$.

Since $D_2$ has a lower power allocation, $R$ has to decode $D_1$ message, then decode $D_2$ message. The outage event at $R$ to not decode $D_2$ message, denoted $\mathcal{A}_{R_2}(\Theta_2)$, is defined as 

\begin{equation}
\mathcal{A}_{R_2}(\Theta_2)\triangleq\frac{P\vert h_{SR}\vert^{2}l_{SR}\,a_2}{\beta P\vert h_{SR}\vert^{2}l_{SR}\,a_1+I_{X_{R}}+I_{Y_{R}}+\sigma^2}< \Theta_2,
\end{equation}
where $\Theta_2=2^{2\mathcal{R}_2}-1$, and $\mathcal{R}_2$ is the target data rate of $D_2$.

Similarly, the outage event at $D_1$ to not decode its intended message in the first phase ($S \rightarrow D_1$), denoted $\mathcal{B}_{D_{1\rightarrow 1}}(\Theta_1)$, is given by

\begin{equation}
\mathcal{B}_{D_{1\rightarrow 1}}(\Theta_1)\triangleq\frac{P\vert h_{SD_1}\vert^{2}l_{SD_1}\,a_1}{P\vert h_{SD_1}\vert^{2}l_{SD_1}a_2+I_{X_{D_1}}+I_{Y_{D_1}}+\sigma^2} < \Theta_1.
\end{equation}
Finally, in order for $D_2$ to decode its intended message, it has to decode $D_1$ message. The outage event at $D_2$ to not decode $D_1$ message in the first phase ($S \rightarrow D_2$), denoted $\mathcal{B}_{D_{2\rightarrow 1}}(\Theta_1)$, and the outage event at $D_2$ to not decode its intended message, denoted $\mathcal{B}_{D_{2\rightarrow 2}}(\Theta_2)$, are respectively given by
\begin{equation}
\mathcal{B}_{D_{2\rightarrow 1}}(\Theta_1)\triangleq\frac{P\vert h_{SD_2}\vert^{2}l_{SD_2}\,a_1}{P\vert h_{SD_2}\vert^{2}l_{SD_2}a_2+I_{X_{D_2}}+I_{Y_{D_2}}+\sigma^2}< \Theta_1,
\end{equation}
and
\begin{equation}
\mathcal{B}_{D_{2\rightarrow 2}}(\Theta_2)\triangleq\frac{P\vert h_{SD_2}\vert^{2}l_{SD_2}\,a_2}{\beta P\vert h_{SD_2}\vert^{2}l_{SD_2}\,a_1+I_{X_{D_2}}+I_{Y_{D_2}}+\sigma^2}< \Theta_2.
\end{equation}
During the second phase, $D_1$ adds the power received from $S$ and from $R$. Hence, the outage event at $D_1$ to not decode its message in the second phase, denoted $\mathcal{C}_{D_{1\rightarrow 1}}(\Theta_1)$, is expressed as
\begin{equation}
\mathcal{C}_{D_{1\rightarrow 1}}(\Theta_1)\triangleq \frac{P\sum_{[SD_1,RD_1]}(\vert h\vert^{2},l)\,a_1}{P\sum_{[SD_1,RD_1]}(\vert h\vert^{2},l)\,a_2+I_{X_{D_1}}+I_{Y_{D_1}}+\sigma^2} < \Theta_1,
\end{equation}
where $$\sum_{[SD_i,RD_i]}(\vert h\vert^{2},l)=\vert h_{SD_i}\vert^{2}l_{SD_i}+\vert h_{RD_i}\vert^{2}l_{RD_2}.$$

In the same way, in the second phase, $D_2$ adds the power received from $S$ and from $R$. Hence, the outage event at $D_2$ to not decode $D_1$ message, denoted $\mathcal{C}_{D_{2\rightarrow1}}(\Theta_1)$, and the outage event at $D_2$ to not decode its message, denoted $\mathcal{C}_{D_{2\rightarrow2}}(\Theta_2)$, are respectively expressed as
\begin{equation}
\mathcal{C}_{D_{2\rightarrow 1}}(\Theta_1)\triangleq \frac{P\sum_{[SD_2,RD_2]}(\vert h\vert^{2},l)\,a_1}{P\sum_{[SD_2,RD_2]}(\vert h\vert^{2},l)\,a_2+I_{X_{D_2}}+I_{Y_{D_2}}+\sigma^2} < \Theta_1,
\end{equation}
and
\begin{equation}
\mathcal{C}_{D_{2\rightarrow 2}}(\Theta_2)\triangleq \frac{P\sum_{[SD_2,RD_2]}(\vert h\vert^{2},l)\,a_2}{\beta P\sum_{[SD_2,RD_2]}(\vert h\vert^{2},l)\,a_1+I_{X_{D_2}}+I_{Y_{D_2}}+\sigma^2} < \Theta_2.
\end{equation}

The overall outage event related to $D_1$, denoted $\textit{O}_{(1)}$, is given by 
\begin{equation}
\textit{O}_{(1)}\triangleq  \Big[ \mathcal{B}_{D_{1\rightarrow 1}}(\Theta_1) \cap \mathcal{A}_{R_1}(\Theta_1) \Big]\cup\Big[\mathcal{A}_{R_1}^C(\Theta_1) \cap \mathcal{C}_{D_{1\rightarrow 1}}(\Theta_1) \Big].  
\end{equation}

Finally, the overall outage event related to $D_2$, denoted $\textit{O}_{(2)}$, is given by 
\begin{align}
\textit{O}_{(2)}\triangleq &\left[\Bigg\{\bigcup_{i=1}^{2} \mathcal{B}_{D_{2\rightarrow    i}}(\Theta_i)\Bigg\} \cap \Bigg\{\bigcup_{i=1}^{2} \mathcal{A}_{R_i}(\Theta_i)\Bigg\}\right] 
\nonumber\\&\bigcup \left[\Bigg\{\bigcap_{i=1}^{2} \mathcal{A}_{R_{i}}^C(\Theta_i)\Bigg\} \cap \Bigg\{\bigcup_{i=1}^{2} \mathcal{C}_{D_{2\rightarrow  i}}(\Theta_i)\Bigg\}\right].  
\end{align}

\subsection{Outage Probability Expressions}

In the following, we will express the outage probability $\textit{O}_{(1)}$ and $\textit{O}_{(2)}$. The probability $\mathbb{P}(\textit{O}_{(1)})$, when $\Theta_1 < a_1/ a_2$, is given by
\begin{align}
       &\mathbb{P}(\textit{O}_{(1)})=1- \mathcal{J}_{(D_1)}\big(\frac{G_{1}}{l_{SD_1}}\big)-
       \mathcal{J}_{(R)}\big(\frac{G_{1}}{l_{SR}}\big)+
       \mathcal{J}_{(D_1)}\big(\frac{G_{1}}{l_{SD_1}}\big)\mathcal{J}_{(R)}\big(\frac{G_{1}}{l_{SR}}\big)
       \nonumber\\
       &+\mathcal{J}_{(R)}\big(\frac{G_{1}}{l_{SR}}\big)-
       \frac{l_{RD_1}\mathcal{J}_{(R)}\big(\frac{G_{1}}{l_{SR}}\big)\mathcal{J}_{(D_1)}\big(\frac{G_{1}}{l_{RD_1}}\big)-l_{SD_1}\mathcal{J}_{(R)}\big(\frac{G_{1}}{l_{SR}}\big)\mathcal{J}_{(D_1)}\big(\frac{G_{1}}{l_{SD_1}}\big)}{l_{RD_1}-l_{SD_1}},
\end{align}
where $G_{1}=\Theta_1/(a_1-\Theta_1 a_2)$, and $\mathcal{J}_{(M)}\Big(\frac{A}{B}\Big)$ is expressed as  \\
\begin{equation}
    \mathcal{J}_{(M)}\Big(\frac{A}{B}\Big)=\mathcal{L}_{I_{X_{M}}}\Big(\frac{A}{B}\Big)\mathcal{L}_{I_{Y_{M}}}\Big(\frac{A}{B}\Big)\exp\Big(-\frac{\sigma^2 A}{P B}\Big).
\end{equation}
The  probability $ \mathbb{P}(\textit{O}_{(2)})$, when $\Theta_1 < a_1/ a_2$ and $\Theta_2 < a_2/ \beta a_1$, is given by\\
\begin{align}
       &\mathbb{P}(\textit{O}_{(2)})=1- \mathcal{J}_{(D_2)}\big(\frac{G_{\mathrm{max}}}{l_{SD_2}}\big)-
       \mathcal{J}_{(R)}\big(\frac{G_{\mathrm{max}}}{l_{SR}}\big)+
       \mathcal{J}_{(D_2)}\big(\frac{G_{\mathrm{max}}}{l_{SD_2}}\big)\mathcal{J}_{(R)}\big(\frac{G_{\mathrm{max}}}{l_{SR}}\big)\nonumber\\
       &+\mathcal{J}_{(R)}\big(\frac{G_{\mathrm{max}}}{l_{SR}}\big)-
       \frac{l_{RD_2}\mathcal{J}_{(R)}\big(\frac{G_{\mathrm{max}}}{l_{SR}}\big)\mathcal{J}_{(D_2)}\big(\frac{G_{\mathrm{max}}}{l_{RD_2}}\big)-l_{SD_2}\mathcal{J}_{(R)}\big(\frac{G_{\mathrm{max}}}{l_{SR}}\big)\mathcal{J}_{(D_2)}\big(\frac{G_{\mathrm{max}}}{l_{SD_2}}\big)}{l_{RD_2}-l_{SD_2}},
\end{align}
where $G_{\mathrm{max}}=\mathrm{max}(G_1,G_2)$, and $G_2=\Theta_2 /(a_2-\Theta_2 \beta a_1)$.

\textit{Proof}:  See Appendix A.\hfill $ \blacksquare $ \\

\subsection{NOMA With $K$-Destinations}
\begin{figure}
\centering
\fbox{%
\fbox{%

\begin{tikzpicture}

\tikzstyle{S}=[circle,very thick,draw=black, fill=red, minimum size=.7cm]
\tikzstyle{R}=[circle,very thick,draw=black, fill=orange, minimum size=.7cm]
\tikzstyle{D1}=[circle,very thick,draw=black, fill=blue, minimum size=.7cm]
\tikzstyle{D2}=[circle,very thick,draw=black, fill=cyan, minimum size=.7cm]
\tikzstyle{DK}=[circle,very thick,draw=black, fill=green, minimum size=.7cm]

\node[S,label=below :{$S$}] (S) at (0,0) {};
\node[R,label=below :{$R$}] (R) at (2, -0.1) {};
\node[D1,label=below :{$D_1$}] (D1) at (4, 2) {};
\node[DK,label=below :{$D_K$}] (DK) at (4, -2) {};
\node[D2,label=below :{$D_2$}] (D2) at (4, 0.7) {};
 \node at ($(D2)!.5!(DK)$) {\vdots};

\draw [->, very thick] (S) -- (R); \draw [->, very thick](S) -- (D1) ;\draw [->, very thick](S) -- (D2) ;
\draw [->, very thick] (S) -- (DK);
\node[draw,text width=3cm] at (2,4) {\textbf{First phase}}; 

\end{tikzpicture}
}
\hspace{1cm}
\fbox{%

\begin{tikzpicture}

\tikzstyle{S}=[circle,very thick,draw=black, fill=red, minimum size=.7cm]
\tikzstyle{R}=[circle,very thick,draw=black, fill=orange, minimum size=.7cm]
\tikzstyle{D1}=[circle,very thick,draw=black, fill=blue, minimum size=.7cm]
\tikzstyle{D2}=[circle,very thick,draw=black, fill=cyan, minimum size=.7cm]
\tikzstyle{DK}=[circle,very thick,draw=black, fill=green, minimum size=.7cm]

\node[S,label=below :{$S$}] (S) at (6,0) {};
\node[R,label=below :{$R$}] (R) at (8, 0) {};
\node[D1,label=below :{$D_1$}] (D1) at (10, 2) {};
\node[D2,label=below :{$D_2$}] (D2) at (10, 0.7) {};
\node[DK,label=below :{$D_K$}] (DK) at (10, -2) {};
\node at ($(D2)!.5!(DK)$) {\vdots};

\draw [->, very thick](R) -- (D1); \draw [->, very thick](R) -- (D2);
\draw [->, very thick] (R) -- (DK);
\node[draw,text width=3cm] at (8,4) {\textbf{Second phase}};
\end{tikzpicture}
}
}
\caption{Transmission scheme using MRC and NOMA considering multiple destinations.} \label{fig:M2}
\end{figure}
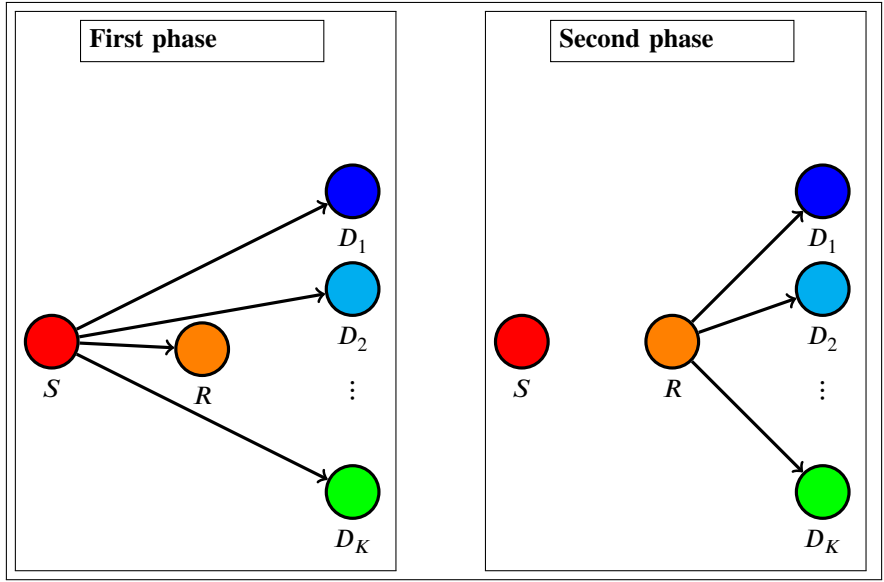

We extend the results of NOMA to $K$-destinations as depicted in Fig.\ref{fig:M2}. 
We generalize the following events to $K$ destination nodes $D_K$ as
\begin{equation}
\mathcal{A}_{R_i}(\Theta_i)\triangleq \frac{P\vert h_{SR}\vert^{2}l_{SR}\,a_i}{P\vert  h_{SR}\vert^{2}l_{SR}\, \big[\beta\sum_{h=1}^{i-1} a_h + \sum_{n=i+1}^{K} a_n \big]+I_{X_{R}}+I_{Y_{R}}+\sigma^2} < \Theta_i,
\end{equation}

\begin{equation}
\mathcal{B}_{D_{i\rightarrow t}}(\Theta_t)\triangleq\frac{P\vert h_{SD_i}\vert^{2}l_{SD_i}\,a_t}{P\vert  h_{SD_i}\vert^{2}l_{SD_i}\, \big[\beta\sum_{h=1}^{t-1} a_h + \sum_{n=t+1}^{K} a_n \big]+I_{X_{D_i}}+I_{Y_{D_i}}+\sigma^2} < \Theta_t,
\end{equation}
and
\begin{align}
&\mathcal{C}_{D_{i\rightarrow t}}(\Theta_1)\triangleq \nonumber\\
&\frac{P\sum_{[SD_i,RD_i]}(\vert h\vert^{2},l)\,a_t}{\big[\beta\sum_{h=1}^{t-1} a_h + \sum_{n=t+1}^{K} a_n \big]P\sum_{[SD_i,RD_i]}(\vert h\vert^{2},l)+I_{X_{D_1}}+I_{Y_{D_1}}+\sigma^2} < \Theta_t.
\end{align}

Note that, when $h>t-1$, then  $\sum_{h=1}^{t-1} a_h=0$, and  when $n>K$, then  $\sum_{n=t+1}^{K} a_n=0$.\\
 
 The outage event at the $i$th destination node, denoted $\textit{O}_{(i)}$, is given by

\begin{align}
\textit{O}_{(i)}\triangleq &\left[\Bigg\{\bigcup_{m=K-i+1}^{K} \mathcal{B}_{D_{i\rightarrow    i-(K-m)}}(\Theta_{i-(K-m)})\Bigg\} \cap \Bigg\{\bigcup_{m=K-i+1}^{K} \mathcal{A}_{R_{i-(K-m)}}(\Theta_{i-(K-m)})\Bigg\}\right] \nonumber\\
\bigcup &\left[\Bigg\{\bigcap_{m=K-i+1}^{K} \mathcal{A}_{R_{i-(K-m)}}^C(\Theta_{i-(K-m)})\Bigg\} \cap \Bigg\{\bigcup_{m=K-i+1}^{K} \mathcal{C}_{D_{i\rightarrow  i-(K-m)}}(\Theta_{i-(K-m)})\Bigg\}\right].\nonumber  
\end{align}

Finally, the outage probability of $D_i$ when $\bigcup\limits_{t=1}^{i} \frac{\,a_t}{\beta \, \sum_{h=1}^{t-1} a_h + \sum_{n=t+1}^{K} a_n}\leq \Theta_{t}$, is expressed by

\begin{align}
       &\mathbb{P}(\textit{O}_{(i)})=\nonumber\\
       &1- \mathcal{J}_{(D_i)}\big(\frac{G_{(i){\textrm{max}}}}{l_{SD_i}}\big)-
       \mathcal{J}_{(R)}\big(\frac{G_{(i){\textrm{max}}}}{l_{SR}}\big)+
       \mathcal{J}_{(D_i)}\big(\frac{G_{(i){\textrm{max}}}}{l_{SD_i}}\big)\mathcal{J}_{(R)}\big(\frac{G_{\mathrm{max}}}{l_{SR}}\big)
       +\mathcal{J}_{(R)}\big(\frac{G_{(i){\textrm{max}}}}{l_{SR}}\big)\nonumber\\
       &-
       \frac{l_{RD_i}\mathcal{J}_{(R)}\big(\frac{G_{(i){\textrm{max}}}}{l_{SR}}\big)\mathcal{J}_{(D_i)}\big(\frac{G_{(i){\textrm{max}}}}{l_{RD_i}}\big)-l_{SD_i}\mathcal{J}_{(R)}\big(\frac{G_{(i){\textrm{max}}}}{l_{SR}}\big)\mathcal{J}_{(D_i)}\big(\frac{G_{(i){\textrm{max}}}}{l_{SD_i}}\big)}{l_{RD_i}-l_{SD_i}},\nonumber
\end{align}

and $G_{(i)\textrm{max}}$ is given by
\begin{align}\label{eq:23}
G_{(i){\textrm{max}}}=&\textrm{max} \Bigg\{ \frac{ \Theta_{i-(K-1)}}{a_{i-(K-1)}- \Theta_{i-(K-1)}[\beta \, \sum_{h=1}^{{i-(K-1)-1}} a_h + \sum_{n=i-(K-1)+1}^{K} a_n]},\nonumber \\
&\quad\quad\quad\frac{ \Theta_{i-(K-2)}}{a_{i-(K-2)}- \Theta_{i-(K-2)}[\beta \, \sum_{h=1}^{{i-(K-2)-1}} a_h + \sum_{n=i-(K-2)+1}^{K} a_n]},...,\nonumber\\
&\quad\quad\quad\frac{ \Theta_{i-(K-l)}}{a_{i-(K-l)}- \Theta_{i-(K-l)}[\beta \, \sum_{h=1}^{{i-(K-l)-1}} a_h + \sum_{n=i-(K-l)+1}^{K} a_n]} \Bigg\},\nonumber
\end{align}
where $l\in\{1,2,...,K\}$, $\Theta_t=2^{2\mathcal{R}_t}-1$, and $\mathcal{R}_t$ is target data rate of $D_t$. We impose the condition that $l>K-i$. 

\section{Laplace Transform Expressions}\label{Section4}

The Laplace transform of the interference originated from the $X$ road at the received node denoted  $M$, is expressed as \cite{J4}

\begin{equation}\label{eq27} 
\mathcal{L}_{I_{X_M}}(s)=\exp\Bigg(-\emph{p}\lambda_{X}\int_\mathbb{R}\dfrac{1}{1+\big(A\Vert \textit{x}-M \Vert^\alpha\big)/sP}dx\Bigg),
\end{equation}
where
\begin{equation}\label{eq28} 
\Vert \textit{x}-M \Vert=\sqrt{\big(m\sin(\theta_{M})\big)^2+\big(x-m \cos(\theta_M) \big)^2 }.
\end{equation}

The Laplace transform of the interference originated from the $Y$ road is given by

\begin{equation}\label{eq29} 
\mathcal{L}_{I_{Y_M}}(s)=\exp\Bigg(-\emph{p}\lambda_{Y}\int_\mathbb{R}\dfrac{1}{1+\big(A\Vert \textit{y}-M \Vert^\alpha\big)/sP}dy\Bigg),
\end{equation}
where
\begin{equation}\label{eq30} 
\Vert \textit{y}-M \Vert=\sqrt{\big(m\cos(\theta_{M})\big)^2+\big(y-m \sin(\theta_M) \big)^2 }.
\end{equation}

The Laplace transform expressions of the interference when $\alpha=2$ are given by

\begin{equation}\label{eq31} 
\mathcal{L}_{I_{X_M}}(s)=\exp\Bigg(-\emph{p}\lambda_{X}\dfrac{sP}{A^{2}}\dfrac{\pi}{\sqrt{\big(m\sin(\theta_{M})\big)^2+sP/A^{2}}}\Bigg),
\end{equation}
and
\begin{equation}\label{eq32} 
\mathcal{L}_{I_{Y_M}}(s)=\exp\Bigg(-\emph{p}\lambda_{Y}\dfrac{sP}{A^{2}}\dfrac{\pi}{\sqrt{\big(m\cos(\theta_{M})\big)^2+sP/A^{2}}}\Bigg).
\end{equation}

\begin{figure}
  \begin{subfigure}{12cm}
    \centering\includegraphics[height=8cm,width=9cm]{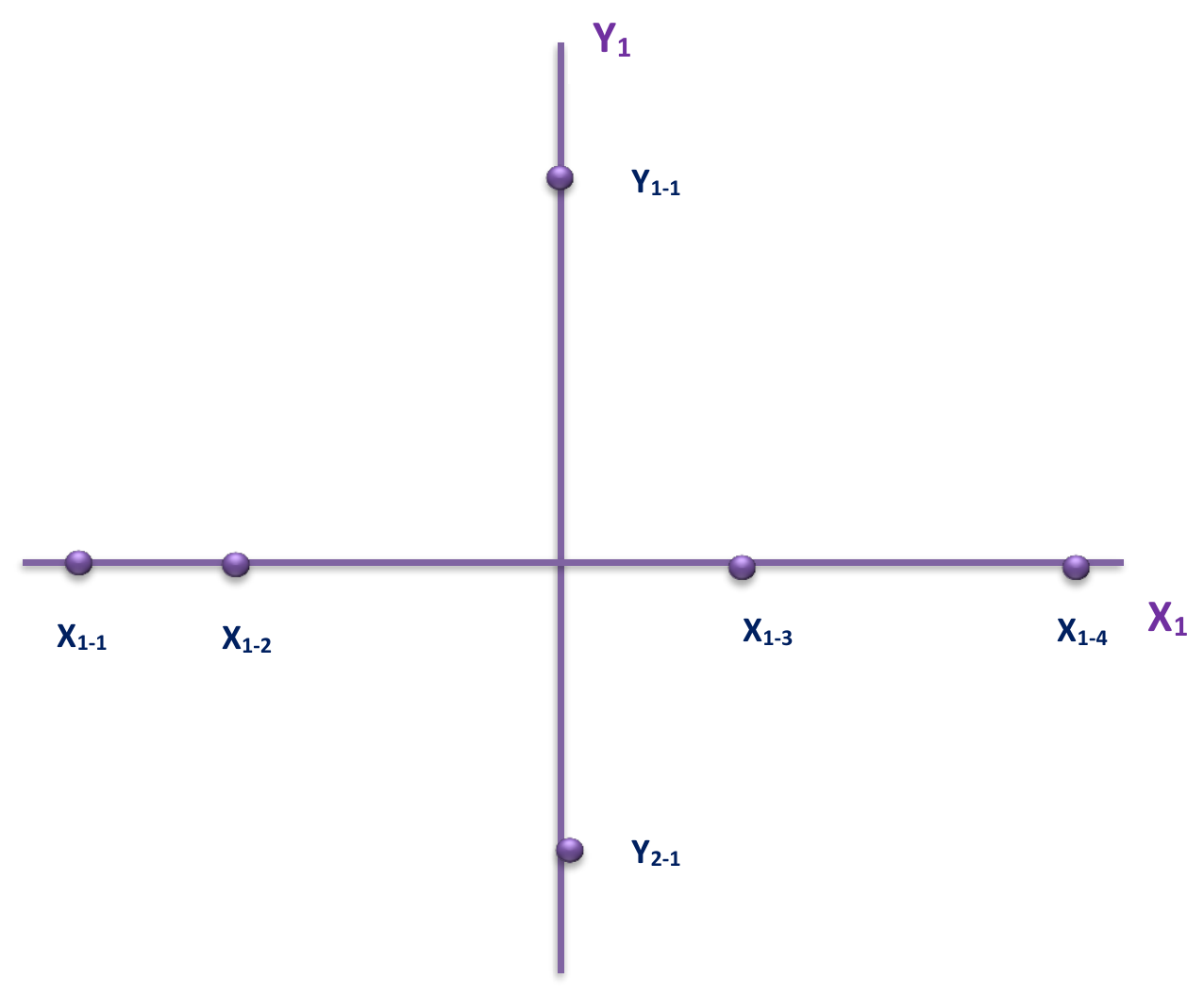}
    \caption{}
     \label{Figurea(a)}
  \end{subfigure}

  \begin{subfigure}{12cm}
    \centering\includegraphics[height=8cm,width=9cm]{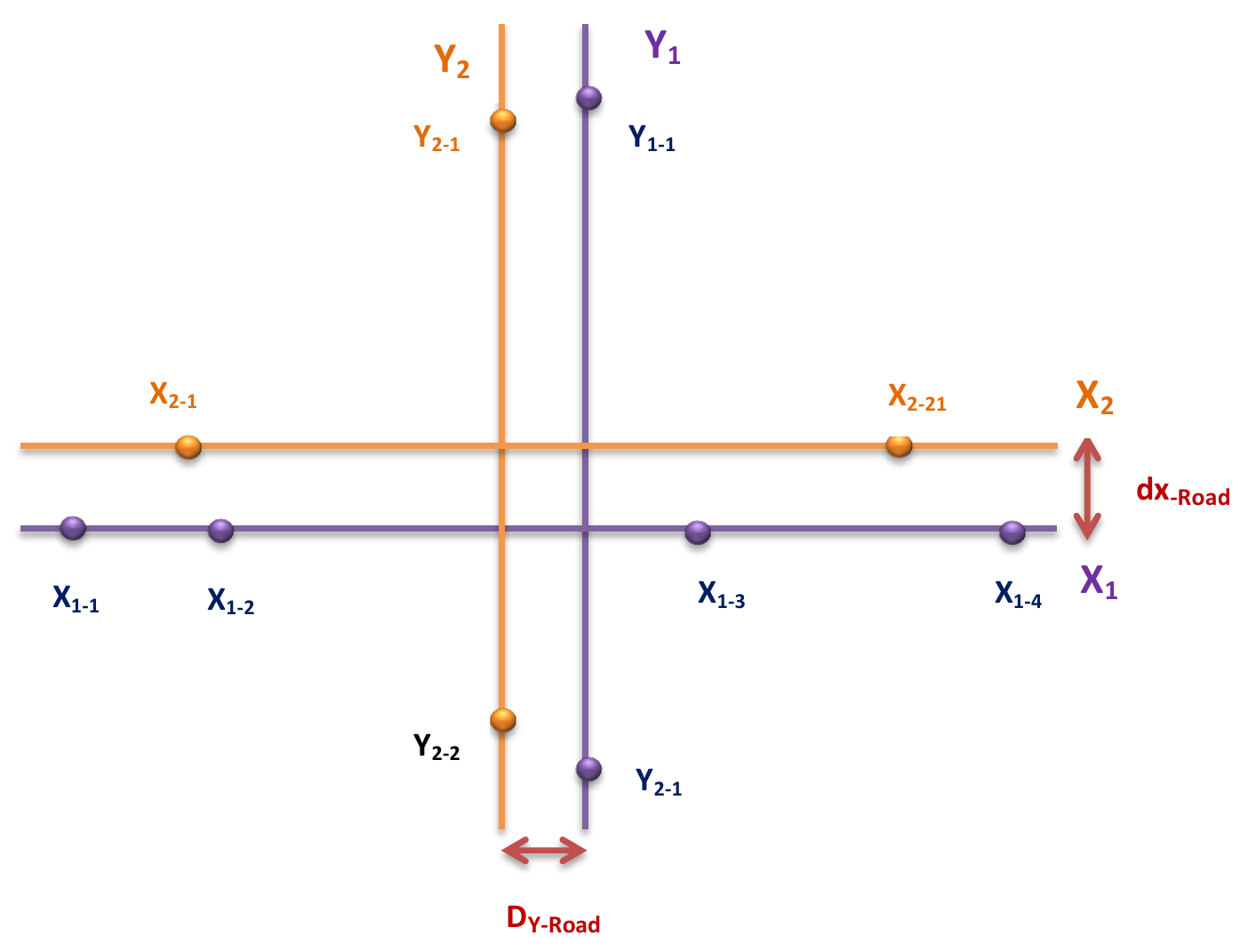}
    \caption{}
         \label{Figureb(b)}
  \end{subfigure}
      \caption{Multiple lane modeling (a) one lane scenario.(b) two lanes scenario.}
      \label{Figure.1}
\end{figure}

\section{Multi Lanes Scenario} \label{Section5}
 Regarding lanes modeling, there are two main approaches to model vehicles on multi-lane roads. The first approach, is the single lane abstraction model or simply the line abstraction model shown in Fig.\ref{Figurea(a)} in which all the traffic lanes are merged into a single lane with the aggregated traffic intensity (see Appendix.C in \cite{rakhshan2016packet}). The second approach is to consider that the traffic is restricted into individual lanes separated by a fixed inter-lane distance, as illustrated in Fig.\ref{Figureb(b)}. 
 We will derive the outage probability for the two road scenario, then generalize the results for multiple lanes.

\subsection{Two-lanes case scenario} 
We address the case where vehicles can drive in two opposite directions, on the horizontal roads and the vertical roads, and further on extend the analysis to $Nb_{lanes}$ number of roads. We refer to the case when we have two roads in the horizontal, and two roads in the vertical as the two-way road case (two lanes on each road). In this case, the horizontal road on which vehicles drive from left to right (resp. right to left) is denoted $X_1$ (resp. $X_2$). The same modification holds for the vertical road on which, vehicles drive from bottom up (resp. top down) is denoted $Y_1$ (resp. $Y_2$). For $\alpha=2$, the expressions of the Laplace transform from the $X_1$ road and the $Y_1$ road at the receiving node $M$ denoted respectively $ \mathcal{L}_{I_{{X_{1}}_M}}(s)$ and $\mathcal{L}_{I_{{Y_{1}}_M}}(s)$, are given by (\ref{49a}) and (\ref{50a}). The expressions of the Laplace transform from the $X_2$ road and from the $Y_2$ road at $M$ are given respectively by
\begin{equation}\label{49a}
\mathcal{L}_{I_{{X_{2}}_M}}(s)=\exp\Bigg(-\emph{p}\lambda_{X_2}\dfrac{s\pi}{\sqrt{(m \sin(\theta_{M})-d_{Y_{Road}})^2+s}}\Bigg),
\end{equation}
and
\begin{equation}\label{50a}
\mathcal{L}_{I_{{Y_{2}}_M}}(s)=\exp\Bigg(-\emph{p}\lambda_{Y_2}\dfrac{s \pi}{\sqrt{(m \cos(\theta_{M})-d_{X_{Road}})^2+s}}\Bigg),
\end{equation}
where $\lambda_{X_2}$ and $\lambda_{Y_2}$ are the intensities of the interferer nodes on the $X_2$ road and $Y_2$ road respectively, and $d_{X_{Road}}$ and $d_{Y_{Road}}$ are distance between $X_1$ and $X_2$, and between $Y_1$ and $Y_2$ respectively. 

\textit{proof}:  See \ref{chapter3:App4}.\hfill $ \blacksquare $\\

In the case when there are two roads on the vertical and two roads on the horizontal, the interference are generated from four roads, the outage probability of $D_1$ and $D_2$ become respectively
\begin{align}
       &\mathbb{P}(\textit{O}_{(1)})=\nonumber\\
       &1- \mathcal{J}^{(2)}_{(D_1)}\big(\frac{G_{1}}{l_{SD_1}}\big)-
       \mathcal{J}^{(2)}_{(R)}\big(\frac{G_{1}}{l_{SR}}\big)+
       \mathcal{J}^{(2)}_{(D_1)}\big(\frac{G_{1}}{l_{SD_1}}\big)\mathcal{J}^{(2)}_{(R)}\big(\frac{G_{1}}{l_{SR}}\big)
       \nonumber\\
       &+\mathcal{J}^{(2)}_{(R)}\big(\frac{G_{1}}{l_{SR}}\big)-
       \frac{l_{RD_1}\mathcal{J}^{(2)}_{(R)}\big(\frac{G_{1}}{l_{SR}}\big)\mathcal{J}^{(2)}_{(D_1)}\big(\frac{G_{1}}{l_{RD_1}}\big)-l_{SD_1}\mathcal{J}^{(2)}_{(R)}\big(\frac{G_{1}}{l_{SR}}\big)\mathcal{J}^{(2)}_{(D_1)}\big(\frac{G_{1}}{l_{SD_1}}\big)}{l_{RD_1}-l_{SD_1}}, \nonumber
\end{align}
and
\begin{align}
       &\mathbb{P}(\textit{O}_{(2)})=\nonumber\\
       &1- \mathcal{J}^{(2)}_{(D_2)}\big(\frac{G_{\mathrm{max}}}{l_{SD_2}}\big)-
       \mathcal{J}^{(2)}_{(R)}\big(\frac{G_{\mathrm{max}}}{l_{SR}}\big)+
       \mathcal{J}^{(2)}_{(D_2)}\big(\frac{G_{\mathrm{max}}}{l_{SD_2}}\big)\mathcal{J}^{(2)}_{(R)}\big(\frac{G_{\mathrm{max}}}{l_{SR}}\big)\nonumber\\
       &+\mathcal{J}^{(2)}_{(R)}\big(\frac{G_{\mathrm{max}}}{l_{SR}}\big)-
       \frac{l_{RD_2}\mathcal{J}^{(2)}_{(R)}\big(\frac{G_{\mathrm{max}}}{l_{SR}}\big)\mathcal{J}^{(2)}_{(D_2)}\big(\frac{G_{\mathrm{max}}}{l_{RD_2}}\big)-l_{SD_2}\mathcal{J}^{(2)}_{(R)}\big(\frac{G_{\mathrm{max}}}{l_{SR}}\big)\mathcal{J}^{(2)}_{(D_2)}\big(\frac{G_{\mathrm{max}}}{l_{SD_2}}\big)}{l_{RD_2}-l_{SD_2}}.\nonumber
\end{align}
where the function is given by
\begin{equation}
    \mathcal{J}^{(2)}_{(M)}\Big(\frac{A}{B}\Big)=\mathcal{L}_{I_{X_{M}}}\Big(\frac{A}{B}\Big)\mathcal{L}_{I_{Y_{M}}}\Big(\frac{A}{B}\Big)\mathcal{L}_{I_{{X_2}_{M}}}\Big(\frac{A}{B}\Big)\mathcal{L}_{I_{{Y_2}_{M}}}\Big(\frac{A}{B}\Big)\exp\Big(-\frac{\sigma^2 A}{P B}\Big)^2.\nonumber
\end{equation}

\subsection{Multi-lanes case scenario}

To generalize the above expressions form $Nb_{lanes}$ roads, we calculate the Laplace transform for the interference for $i^{th} X$ road, and $i^{th} Y$ road when $\alpha=2$ is respectively given by:

\begin{equation}\label{51}
\mathcal{L}_{I_{{X_{i}}_{M}}}(s)=\exp\Bigg(-\emph{p}\lambda_{X_i}\dfrac{s\pi}{\sqrt{(m \sin(\theta_{{M}})-\sum_{i=1}^{Nb_{lanes}-1}(i-1)d_{Y_{Road}})^2+s}}\Bigg)
\end{equation}
\begin{equation}\label{52}
\mathcal{L}_{I_{{Y_{i}}_{M}}}(s)=\exp\Bigg(-\emph{p}\lambda_{Y_i}\dfrac{s\pi}{\sqrt{(m \cos(\theta_{{M}})-\sum_{i=1}^{Nb_{lanes}-1}(i-1)d_{X_{Road}})^2+s}}\Bigg)
\end{equation} 
where $\lambda_{X_i}$ and $\lambda_{Y_i}$ are the intensities of the interferer nodes on the $X_i$ road and $Y_i$ road respectively. Hence the outage probability of $D_1$ and $D_2$ are respectively given by

\begin{align}
       &\mathbb{P}(\textit{O}_{(1)})=\nonumber\\
       &1- \mathcal{J}^{(Nb_{lanes})}_{(D_1)}\big(\frac{G_{1}}{l_{SD_1}}\big)-
       \mathcal{J}^{(Nb_{lanes})}_{(R)}\big(\frac{G_{1}}{l_{SR}}\big)+
       \mathcal{J}^{(Nb_{lanes})}_{(D_1)}\big(\frac{G_{1}}{l_{SD_1}}\big)\mathcal{J}^{(Nb_{lanes})}_{(R)}\big(\frac{G_{1}}{l_{SR}}\big)
       \nonumber\\
       &+\mathcal{J}^{(Nb_{lanes})}_{(R)}\big(\frac{G_{1}}{l_{SR}}\big)\nonumber\\
       &-\frac{l_{RD_1}\mathcal{J}^{(Nb_{lanes})}_{(R)}\big(\frac{G_{1}}{l_{SR}}\big)\mathcal{J}^{(Nb_{lanes})}_{(D_1)}\big(\frac{G_{1}}{l_{RD_1}}\big)-l_{SD_1}\mathcal{J}^{(Nb_{lanes})}_{(R)}\big(\frac{G_{1}}{l_{SR}}\big)\mathcal{J}^{(Nb_{lanes})}_{(D_1)}\big(\frac{G_{1}}{l_{SD_1}}\big)}{l_{RD_1}-l_{SD_1}},
\end{align}
and
\begin{align}
       &\mathbb{P}(\textit{O}_{(2)})=\nonumber\\
       &1- \mathcal{J}^{(Nb_{lanes})}_{(D_2)}\big(\frac{G_{\mathrm{max}}}{l_{SD_2}}\big)-
       \mathcal{J}^{(Nb_{lanes})}_{(R)}\big(\frac{G_{\mathrm{max}}}{l_{SR}}\big)+
       \mathcal{J}^{(Nb_{lanes})}_{(D_2)}\big(\frac{G_{\mathrm{max}}}{l_{SD_2}}\big)\mathcal{J}^{(Nb_{lanes})}_{(R)}\big(\frac{G_{\mathrm{max}}}{l_{SR}}\big)\nonumber\\
       &+\mathcal{J}^{(Nb_{lanes})}_{(R)}\big(\frac{G_{\mathrm{max}}}{l_{SR}}\big)\nonumber\\
       &-
       \frac{l_{RD_2}\mathcal{J}^{(Nb_{lanes})}_{(R)}\big(\frac{G_{\mathrm{max}}}{l_{SR}}\big)\mathcal{J}^{(Nb_{lanes})}_{(D_2)}\big(\frac{G_{\mathrm{max}}}{l_{RD_2}}\big)-l_{SD_2}\mathcal{J}^{(Nb_{lanes})}_{(R)}\big(\frac{G_{\mathrm{max}}}{l_{SR}}\big)\mathcal{J}^{(Nb_{lanes})}_{(D_2)}\big(\frac{G_{\mathrm{max}}}{l_{SD_2}}\big)}{l_{RD_2}-l_{SD_2}}.
\end{align}
where 
\begin{equation}
    \mathcal{J}^{(Nb_{lanes})}_{(M)}\Big(\frac{A}{B}\Big)=\exp\Big(-\frac{\sigma^2 A}{P B}\Big)^{Nb_{lanes}}\times\prod_{i=1}^{{Nb_{lanes}}}\mathcal{L}_{I_{{X_i}_{M}}}\Big(\frac{A}{B}\Big)\mathcal{L}_{I_{{Y_i}_{M}}}\Big(\frac{A}{B}\Big).
\end{equation}

\begin{figure}[]
\centering
\includegraphics[height=8cm,width=9cm]{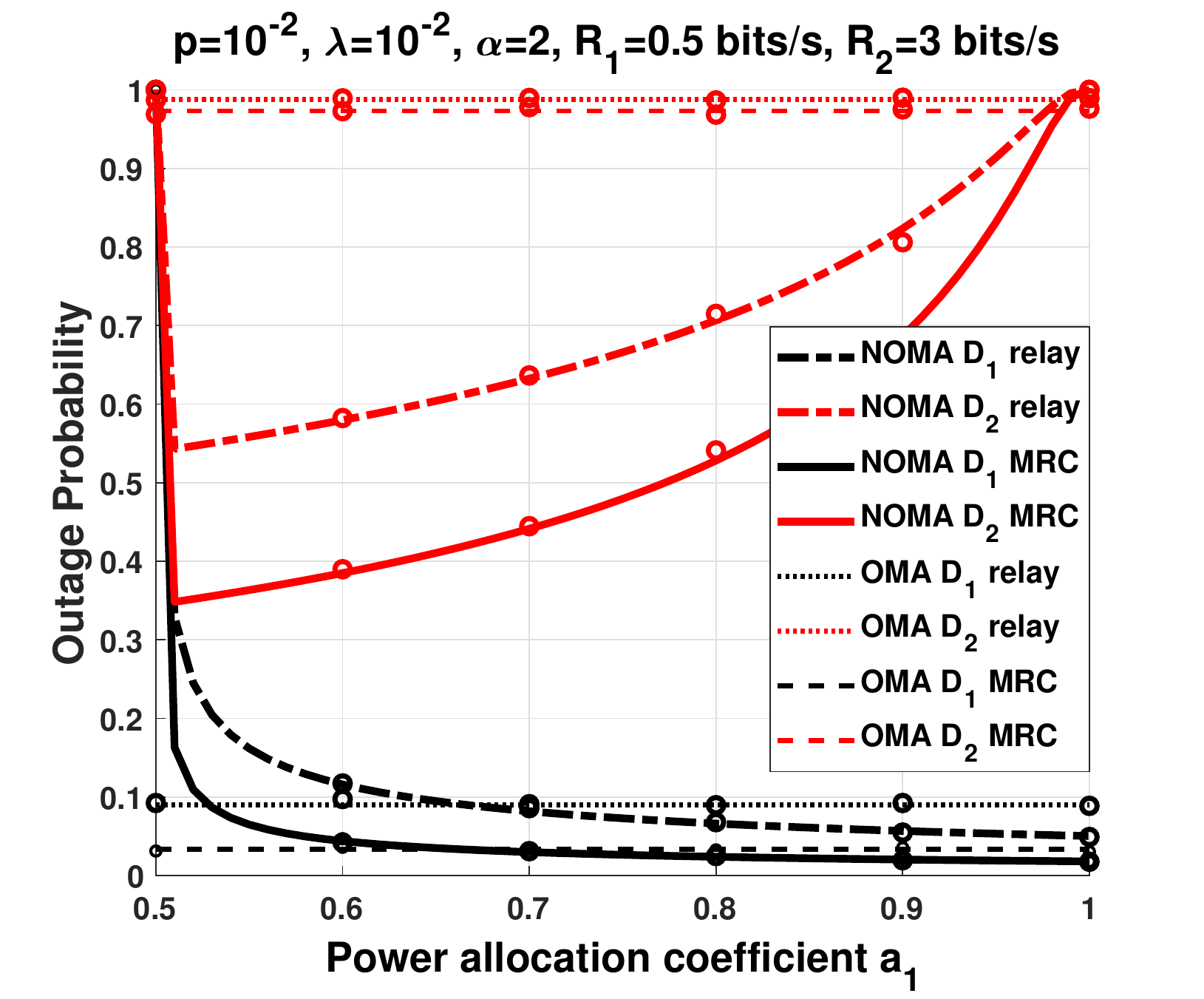}
\caption{Outage probability as a function of $a_1$ considering NOMA and OMA.}
\label{Fig2}
\end{figure}

\section{Simulations and Discussions}\label{Section6}
In this section, we evaluate the performance of MRC with NOMA at road intersections. Monte-Carlo simulation are carried out by generating samples (which correspond to the interfering vehicles) according to a PPP, and we average over $50,000$ iterations of Rayleigh fading channel coefficients. The Monte-Carlo simulations match the theoretical analysis, which confirm the accuracy of our results. 
Unless stated otherwise,  $\beta=0$, $S=[100,0]$, $R=[50,0]$, $D_1=[0,0]$ and $D_2=[0,-10]$. We set, without loss of generality,  $\lambda_X = \lambda_Y = \lambda$. 

\begin{figure}[]
\centering
\includegraphics[height=8cm,width=9cm]{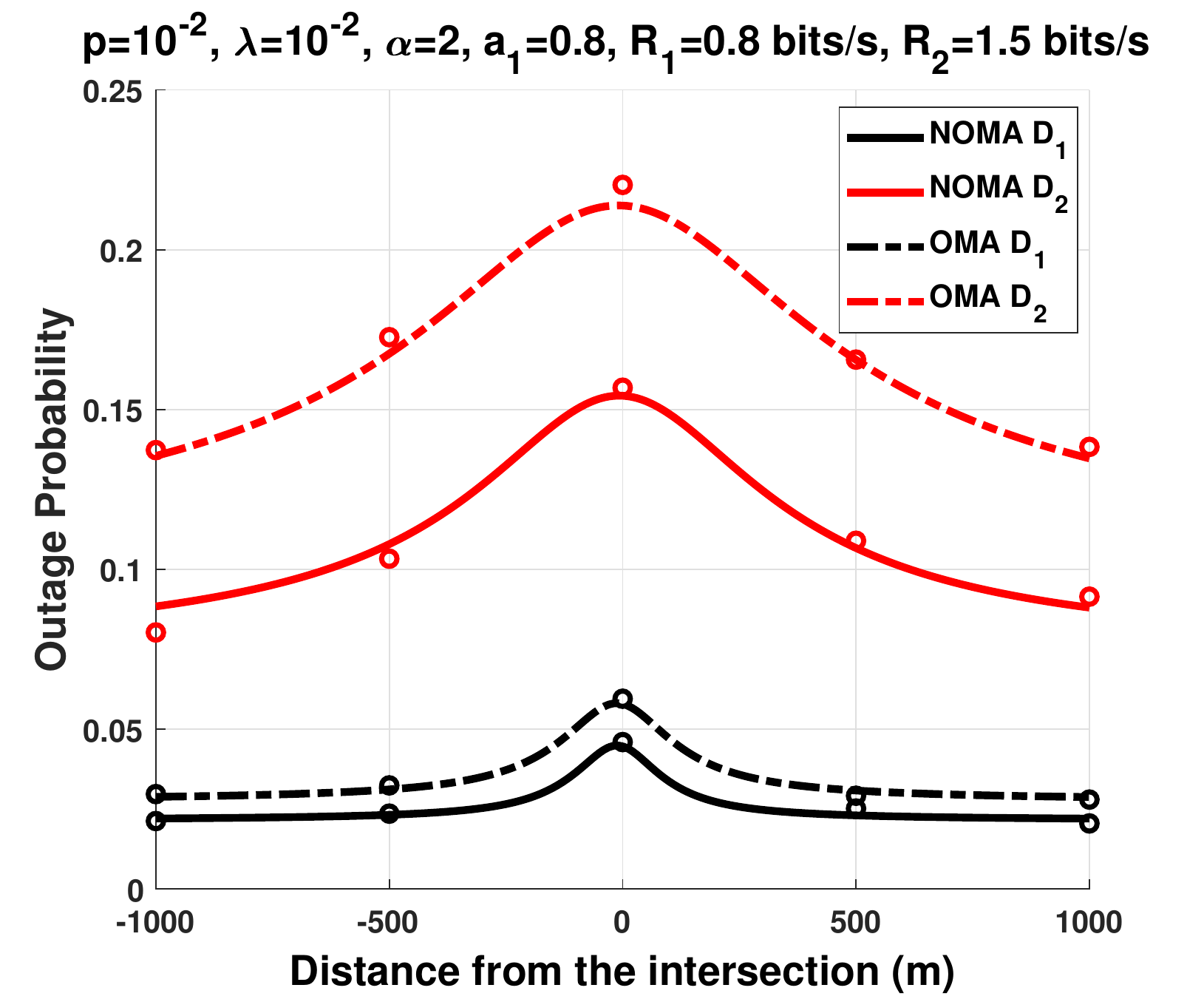}
\caption{Outage probability as a function of the distance from the intersection considering NOMA and OMA.}
\label{Fig3}
\end{figure}
Fig.\ref{Fig2} compares the outage probability as a function of $a_1$, considering a NOMA relay transmission \cite{J4}, NOMA MRC transmission (the proposed method), OMA relay transmission, and OMA MRC transmission \cite{J2}. 
The figure shows that implementing MRC with NOMA offers a significant improvement on the performance compared to the relay transmission. This improvement is event greater for $D_2$. To quantify this improvement, we notice that, MRC with NOMA offers decrease of the outage probability of $34\%$ compared to the relay transmission with NOMA. Whereas the improvement of MRC with OMA is $2\%$ compared to the relay transmission with OMA. We can also notice that there is an improvement of $60\%$ in terms of outage probability when using MRC with NOMA compared to MRC with OMA.

Fig.\ref{Fig3} depicts the outage probability as a function of the distance between the nodes and the intersection. 
We can see, from Fig.\ref{Fig3}, that the outage probability has a peak at the intersection. This can be explained by the fact that the interfering vehicles from both $X$ and $Y$ road contribute to the aggregate interference. Whereas only one road contribute to the aggregate interference when the 
nodes are far from the intersection. We also see that implementing MRC with NOMA offers a better performance than MRC with OMA for $D_1$ and $D_2$. 

\begin{figure}[]
\centering
\includegraphics[height=8cm,width=9cm]{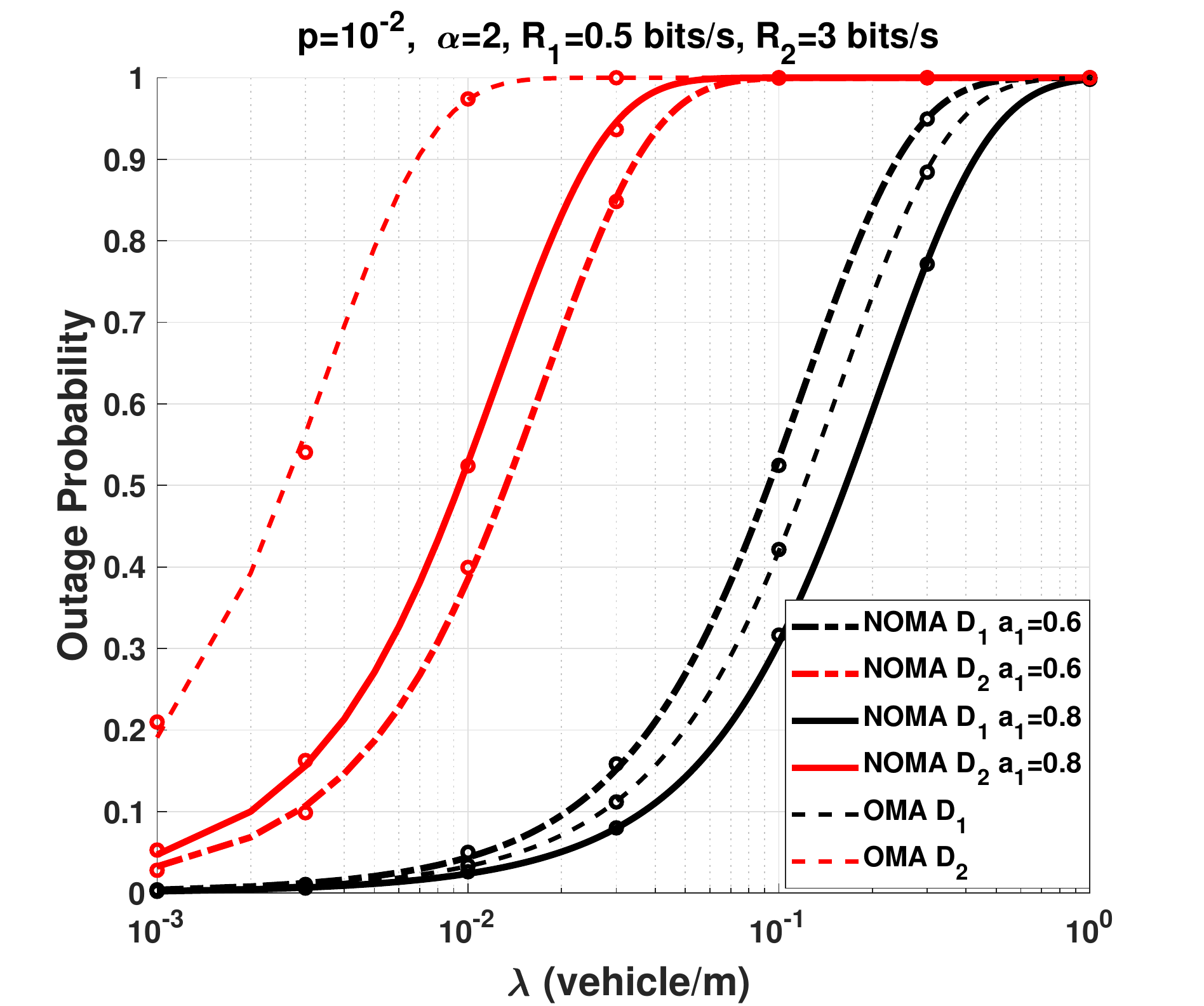}
\caption{Outage probability as a function of $\lambda$ considering NOMA and OMA.}
\label{Fig4}
\end{figure}

Fig.\ref{Fig4} plots the outage probability as a function of the vehicles density $\lambda$.
We notice that as the intensity of the interfering vehicles increases, the outage probability increases.
The reason is that as the number of vehicles increases, the aggregate of interference increases at the receiver
node, which decreases the SIR and increases the outage probability. Note that the value of $a_1$ has to be chosen carefully, since when $a_1=0.6$, MRC with NOMA offers a better performance than MRC with OMA for $D_1$ and $D_2$. which is not the case when $a_1=0.8$

\begin{figure}[]
\centering
\includegraphics[height=8cm,width=9cm]{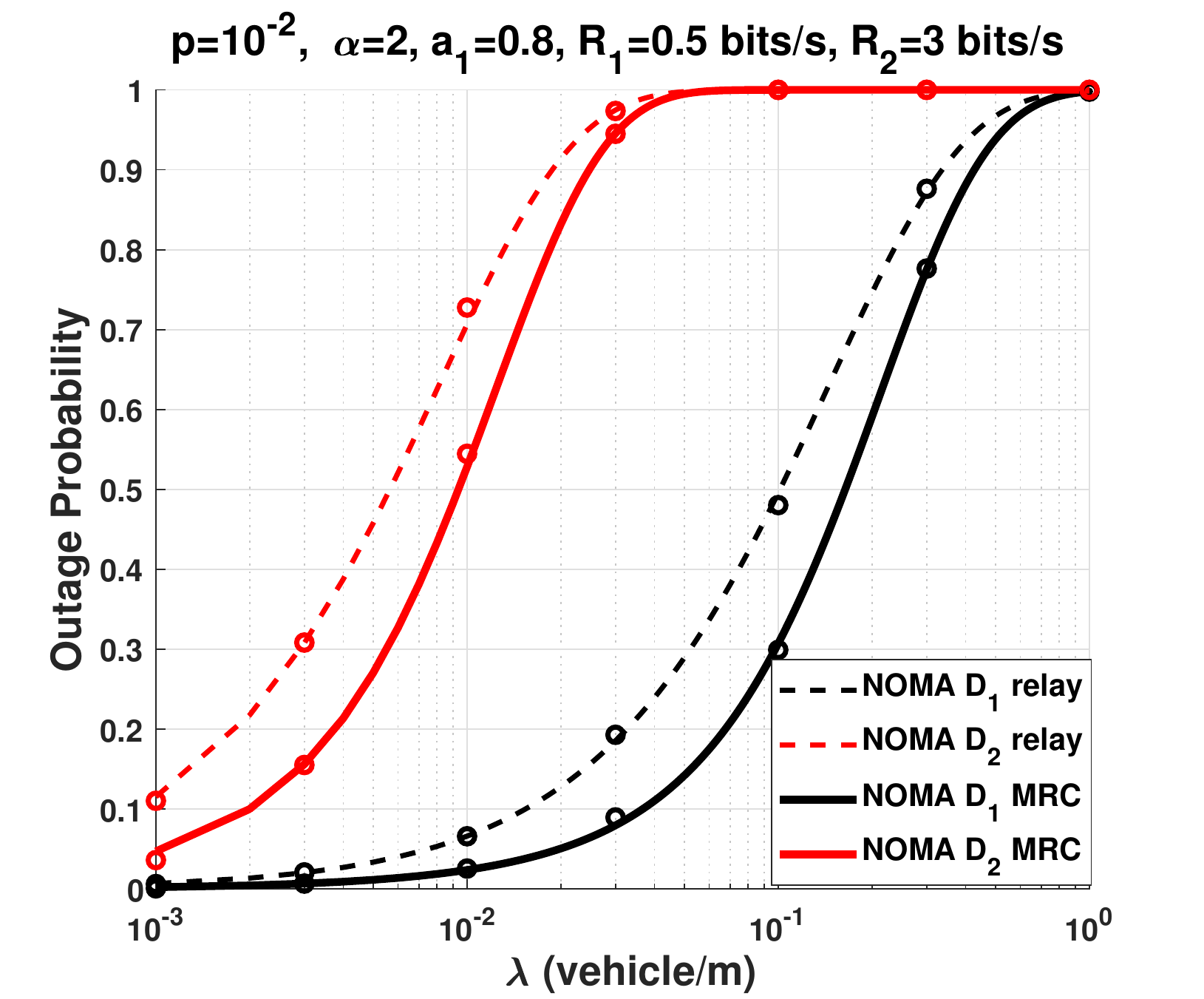}
\caption{Outage probability as a function of $\lambda$ considering NOMA using different transmission schemes. }
\label{Fig5}
\end{figure}

Fig.\ref{Fig5} shows the outage probability as a function of $\lambda$ considering NOMA using different transmission schemes. We can clearly see that the MRC using NOMA outperforms the classical relay transmission using NOMA. This holds true for both $D_1$ and $D_2$. This result is intuitive since in the relay transmission using NOMA, $D_1$ and $D_2$ decode the message transmitted by the relay. However, in the MRC transmission scheme using NOMA, $D_1$ and $D_2$ combine the signal from the source, and from the relay, which increases the power at the $D_1$ and $D_2$, and consequently increases the SINR.

\begin{figure}
  \begin{subfigure}{12cm}
    \centering\includegraphics[height=8cm,width=9cm]{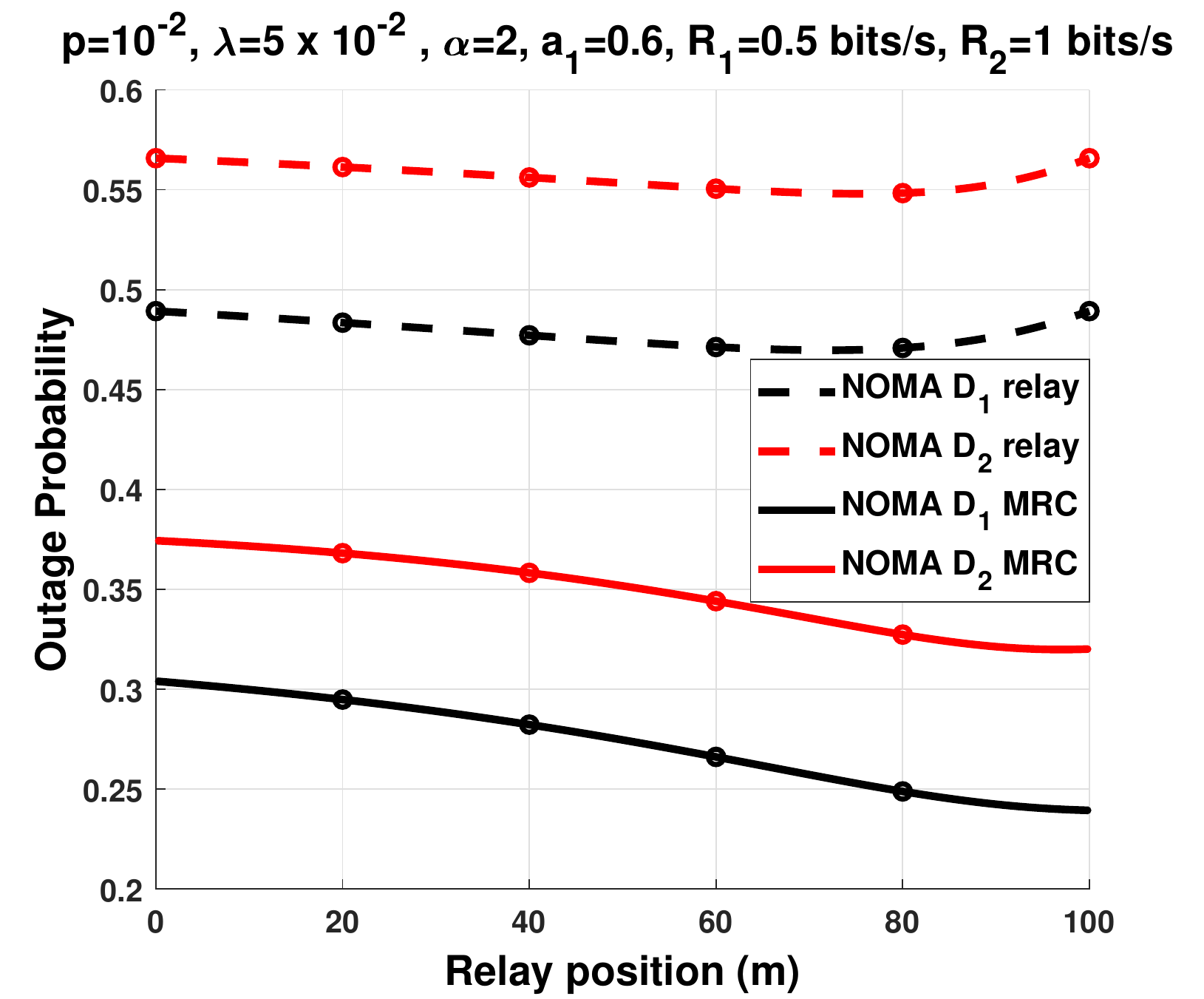}
    \caption{}
     \label{Fig6a}
  \end{subfigure}

  \begin{subfigure}{12cm}
    \centering\includegraphics[height=8cm,width=9cm]{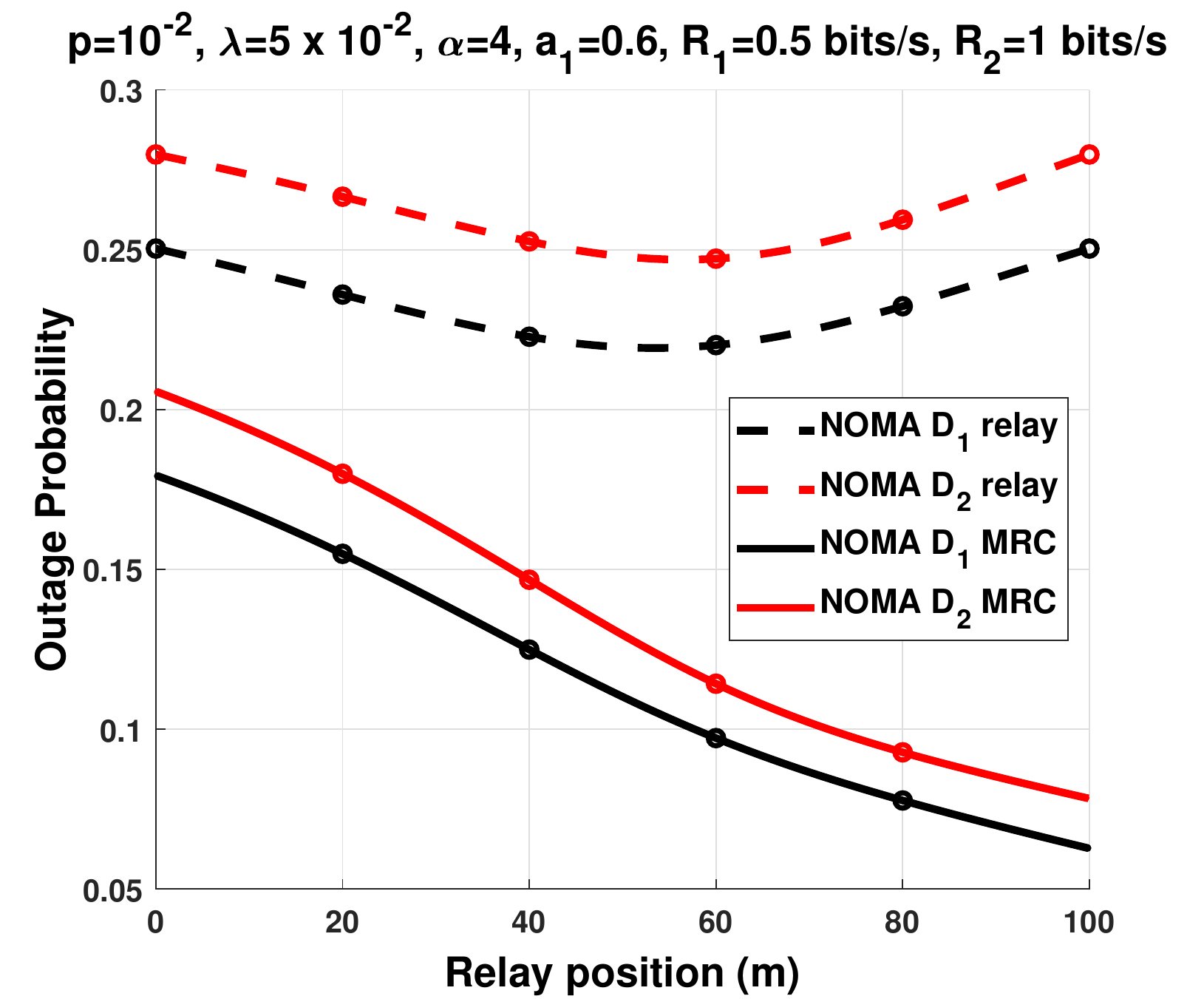}
    \caption{}
         \label{Fig6b}
  \end{subfigure}
      \caption{Outage probability as a function of the relay position. (a) $\alpha=2$. (a) $\alpha=4$.}
      \label{Fig6}
\end{figure}

Fig.\ref{Fig6} depicts the outage probability as a function of the relay position, 
using a relay transmission and MRC transmission considering NOMA. We set, without loss of generality, 
$\Vert S-D_1 \Vert = \Vert S-D_2 \Vert=100$m.

We can notice from Fig.\ref{Fig6a} that when $\alpha=2$, the optimal position for the relay using 
a relay transmission is near the destinations, $D_1$ and $D_2$, whereas for MRC, the optimal relay 
position is when the relay is close to the destination nodes.

When $\alpha=4$, we can see, from Fig.\ref{Fig6b}, that the best position for the relay is at mid-distance
between $S$ and the destination nodes when using the relay transmission. But, when using MRC, the best relay position is when the relay is near the destination nodes.
This is because, when the relay is near the destination, the channel coefficients between and $S$
and the destination, and between $R$ and the destination are decorrelated, which increases the diversity gain.

\begin{figure}[]
\centering
\includegraphics[height=8cm,width=9cm]{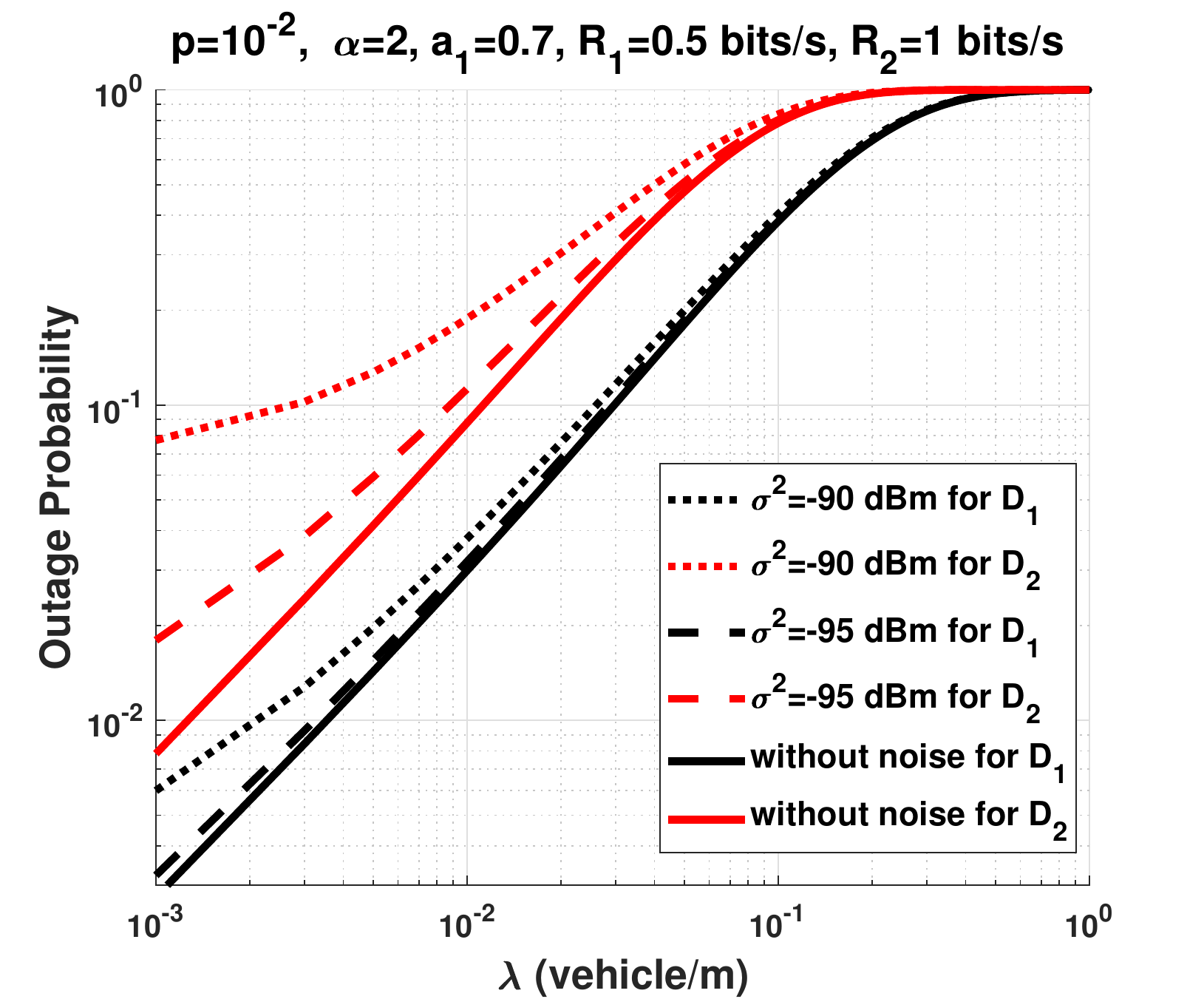}
\caption{Outage probability as a function of $\lambda$ for several noise power values.}
\label{Fig8}
\end{figure}

We can see form the Fig.\ref{Fig8} that the noise power greatly impact the performance 
only for low values of $\lambda$. However, as the value of $\lambda$ increases,
the performance when considering noise power and without noise power tends the 
same values. This because for high value of $\lambda$, the power of noise
become negligible compared to the power of interference.

\begin{figure}
  \begin{subfigure}{12cm}
    \centering\includegraphics[height=8cm,width=9cm]{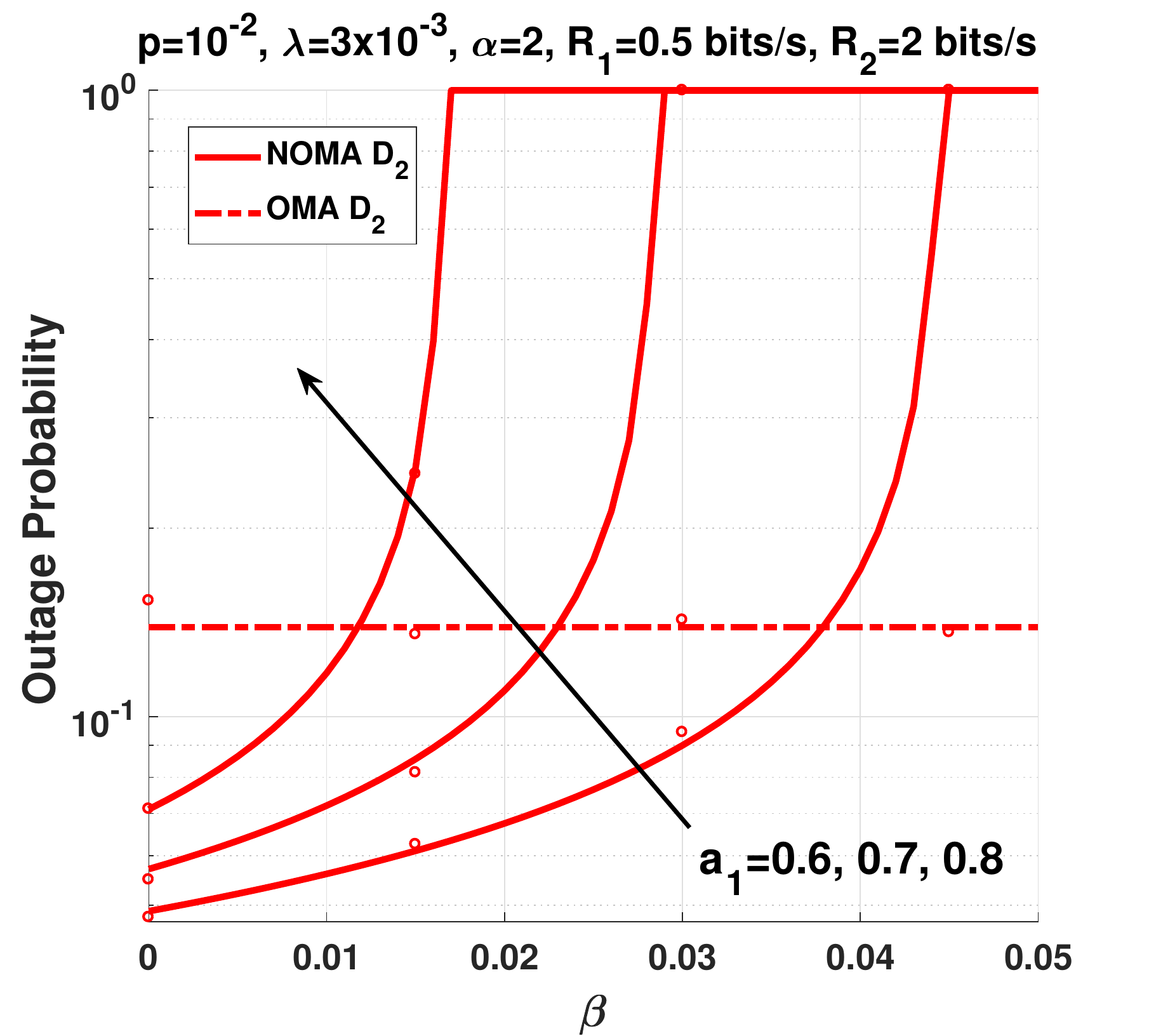}
    \caption{}
     \label{Fig7a}
  \end{subfigure}

  \begin{subfigure}{12cm}
    \centering\includegraphics[height=8cm,width=9cm]{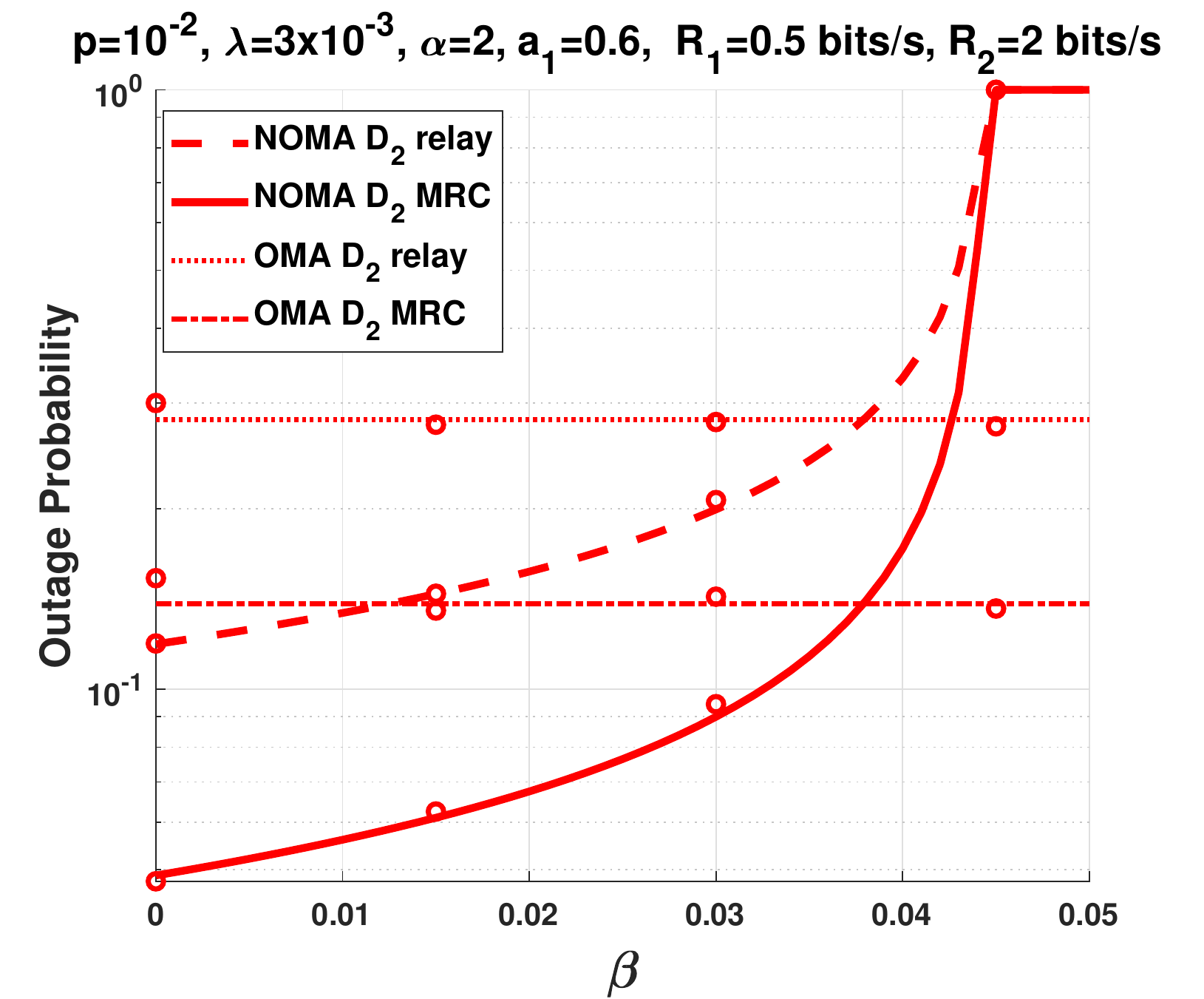}
    \caption{}
         \label{Fig7b}
  \end{subfigure}
      \caption{Outage probability of $D_2$ as a function of $\beta$ considering NOMA and OMA. (a) NOMA and OMA considering MRC transmission. (b) NOMA and OMA considering MRC transmission and relay transmission.}
      \label{Fig7}
\end{figure}

Fig.\ref{Fig7} shows the impact of $\beta$ on the performance in terms of outage probability.
We can see from Fig.\ref{Fig7a} that for low values of $\beta$ the outage probability considering
NOMA is lower than OMA when using MRC transmission. However, as the value of $\beta$ increases, the outage probability
of NOMA increases. We can also see that as $a_1$ decreases, the values of the effect of $\beta$ becomes less dominant. This because as we allocate more power to $D_2$, it increases the SINR at $D_2$
hence decreasing the outage probability.
We can also see from Fig.\ref{Fig7b} that the MRC outperforms the relay transmission for both NOMA 
and OMA. However, we can see that the value of $\beta$ when OMA outperforms NOMA
is the same for MRC and the relay transmission.

\begin{figure}
  \begin{subfigure}{12cm}
    \centering\includegraphics[height=8cm,width=9cm]{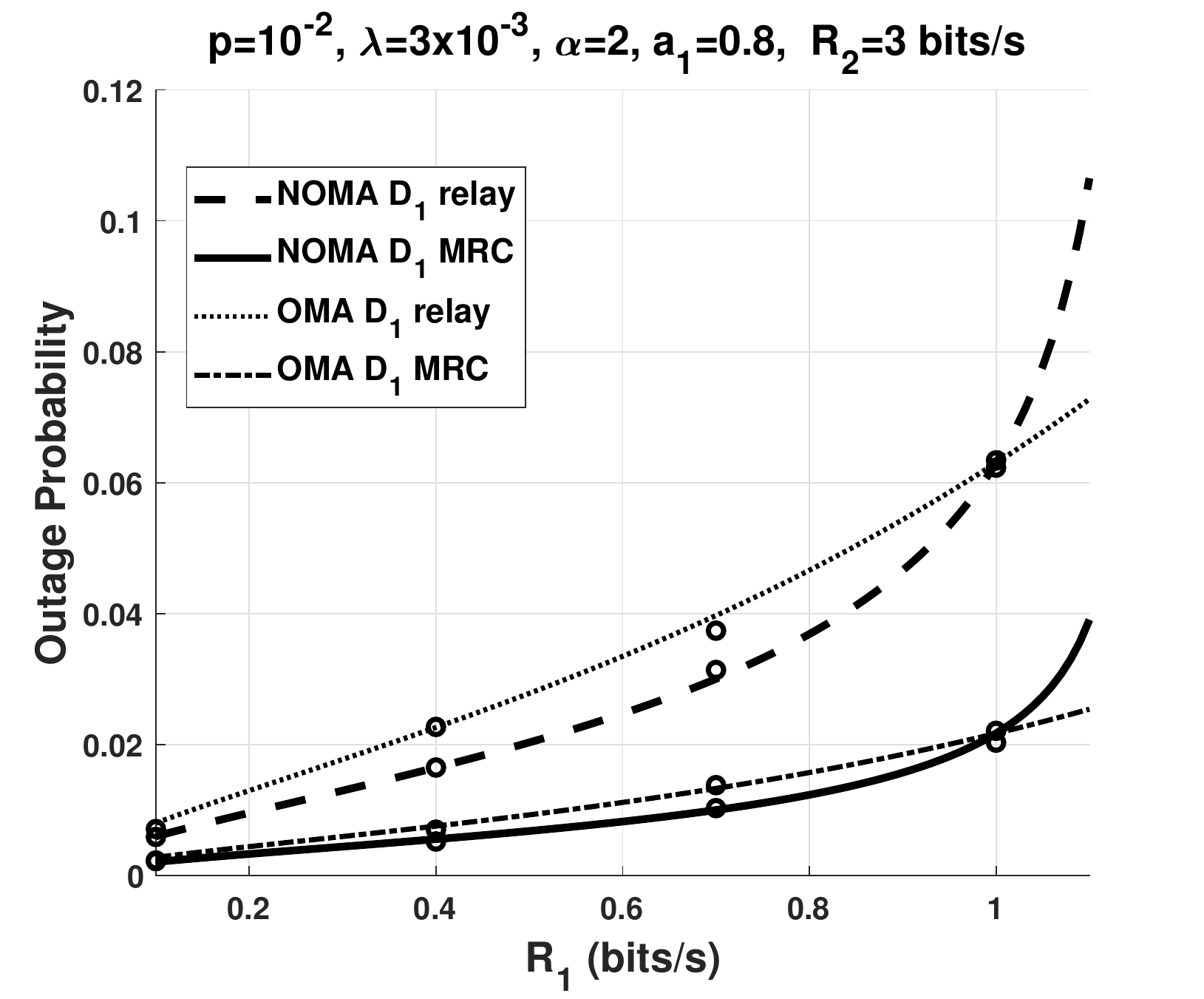}
    \caption{}
     \label{Fig9a}
  \end{subfigure}

  \begin{subfigure}{12cm}
    \centering\includegraphics[height=8cm,width=9cm]{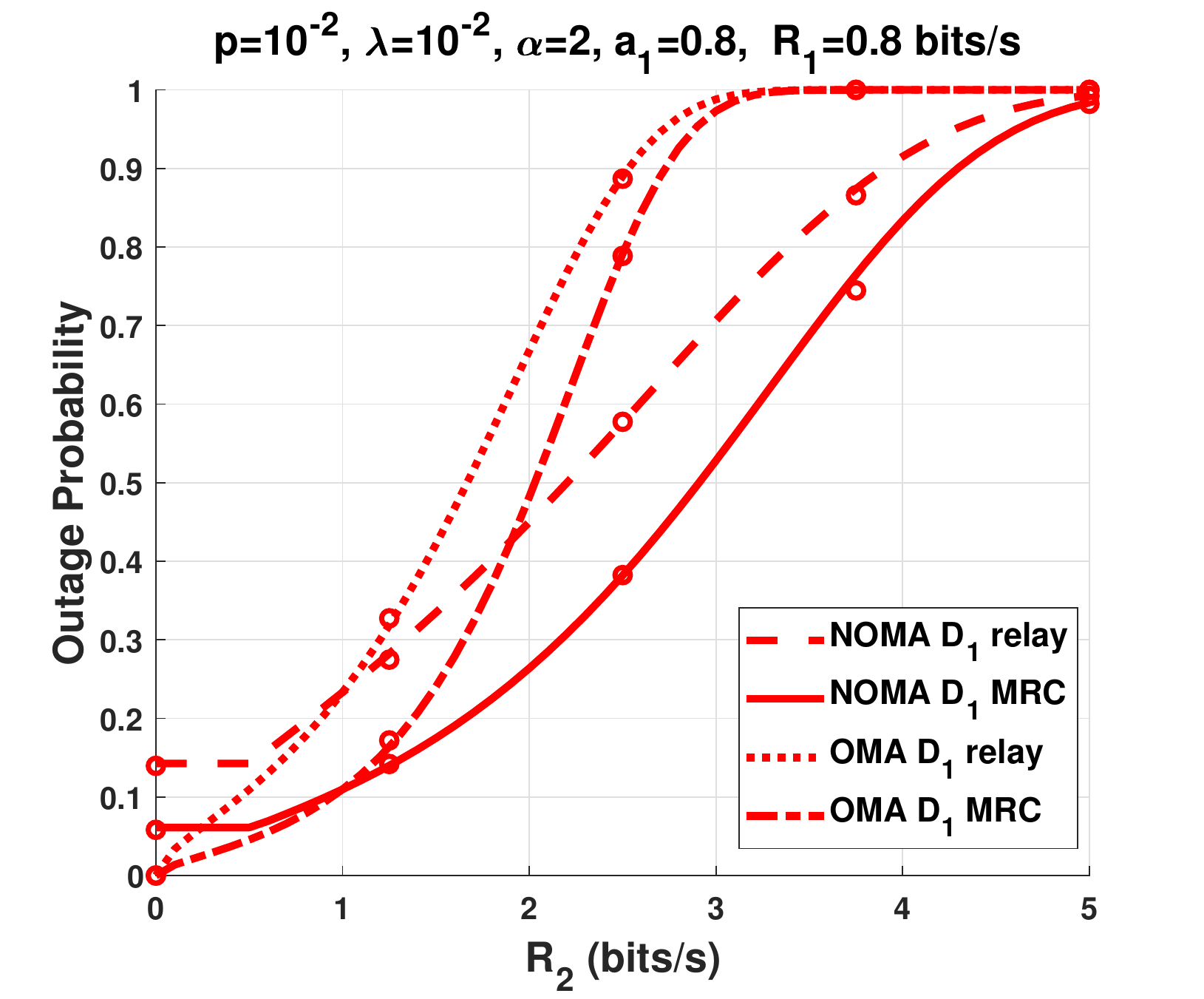}
    \caption{}
         \label{Fig9b}
  \end{subfigure}
      \caption{Outage probability as a function of data rate considering NOMA and OMA. (a) Outage probability as a function of $\mathcal{R}_1$. (b) Outage probability as a function of $\mathcal{R}_2$.}
      \label{}
\end{figure}

\begin{figure}
  \begin{subfigure}{12cm}
    \centering\includegraphics[height=8cm,width=9cm]{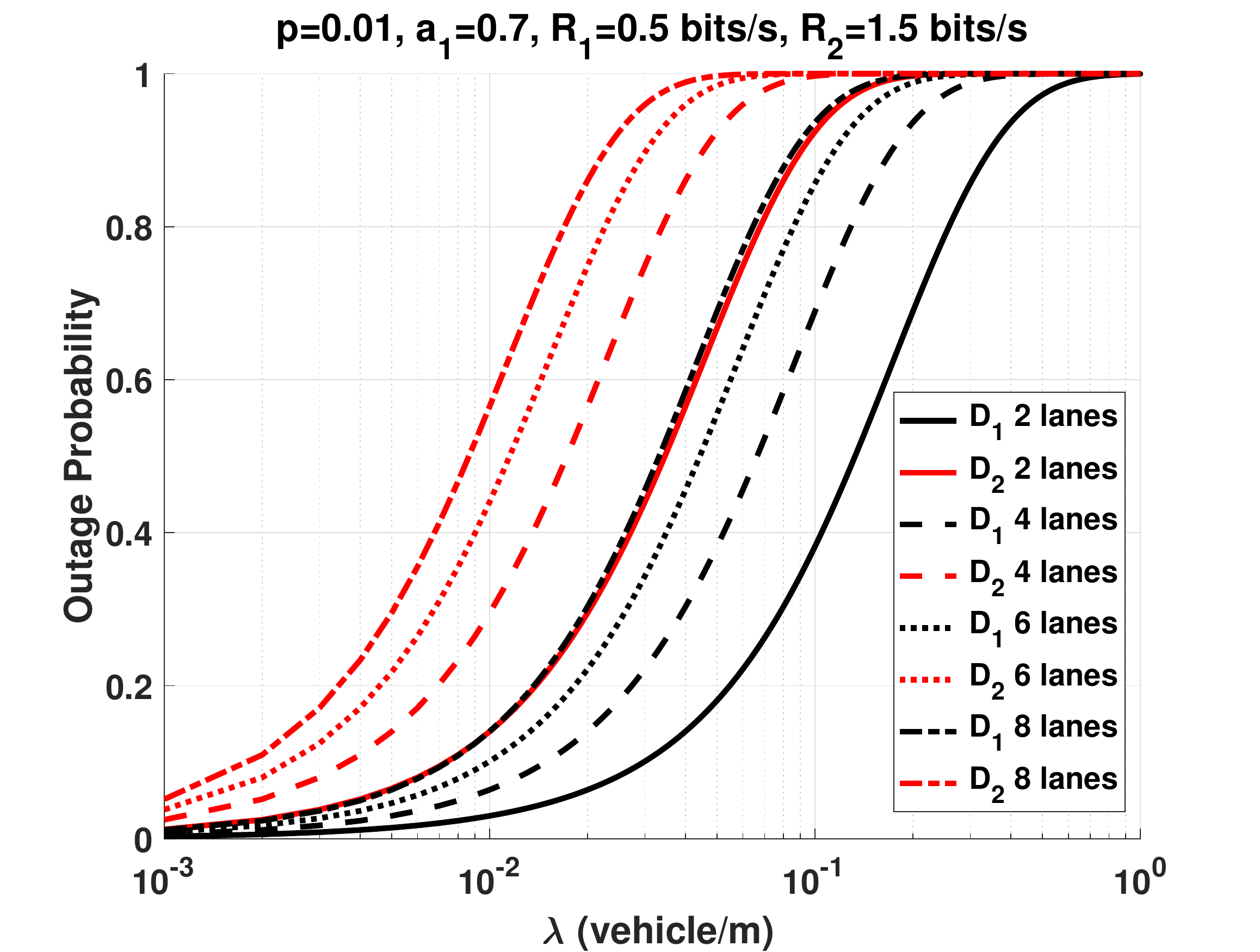}
    \caption{}
     \label{Figure13(a)}
  \end{subfigure}

  \begin{subfigure}{12cm}
    \centering\includegraphics[height=8cm,width=9cm]{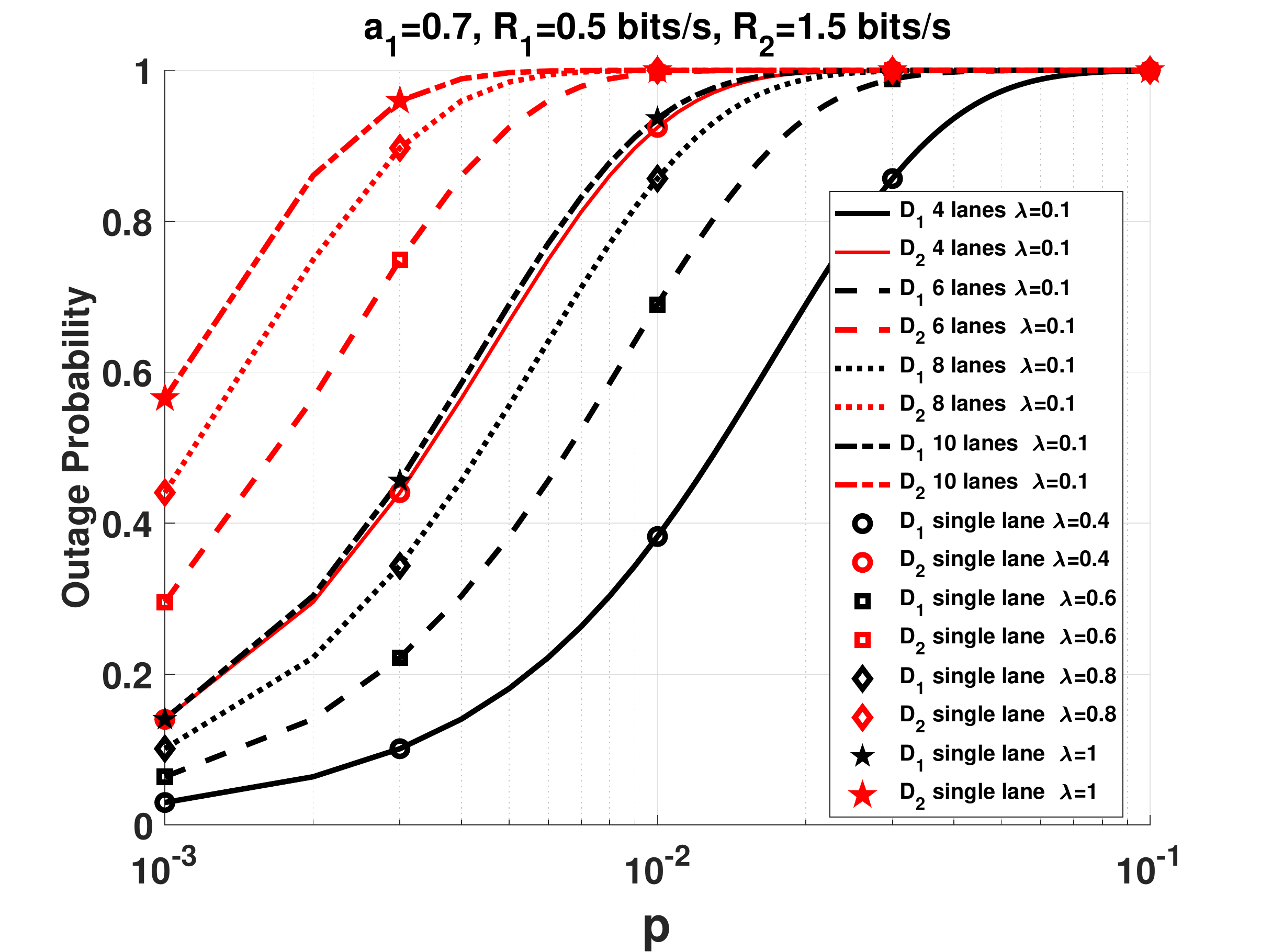}
    \caption{}
    \label{Figure13(b)}
  \end{subfigure}
      \caption{ Multiple lanes modeling considering MRC NOMA. (a) outage probability as a function of $\lambda$ for $Nb_{lane}=\{2,4,6,8\}$. (b) outage probability as a function of $p$ for the single lane model and the multiple lane model for several values of $Nb_{lane}$. }
      \label{Figure.13}
\end{figure}

Finally, we investigate the impact of the data rates $\mathcal{R}_1$ and $\mathcal{R}_2$
on the performance considering NOMA and OMA using MRC and the relay transmission. We can see from Fig.\ref{Fig9a} that as $\mathcal{R}_1$ increases, the outage probability of $D_1$
increases. This is intuitive since increasing the data rate increases the decoding threshold
which increases the outage probability. We can also see that NOMA offers better performance than OMA. However, as $\mathcal{R}_1$ increases, OMA outperforms NOMA for both MRC transmission and relay transmission.

Also, we can see from Fig.\ref{Fig9b} that from small values of $\mathcal{R}_2$, that is, $\mathcal{R}_2<0.5$ bit/s, OMA offers better performance than NOMA in terms of outage probability. This is because, unlike the vehicle $D_1$, the vehicle $D_2$ has to decode $D_1$ message first, and then decode its own message. Hence, $\mathbb{P}(D_2)$ depends solely on $\mathcal{R}_1$ for small values of $\mathcal{R}_2$.
 We also notice that, for large values of $\mathcal{R}_2$ $(\mathcal{R}_2>2 \textrm{bit/s})$, NOMA has better performance in terms of outage probability than OMA. This because for large values of $\mathcal{R}_2$, the decoding threshold of OMA increases linearly since it is multiplied by a factor of 4.  This proves that cooperative NOMA has a better outage performance for high data rates. Finally, we can see that MRC transmission outperforms cooperative transmission for both NOMA and OMA.

Fig.\ref{Figure13(a)} plots NOMA outage probability as a function of $\lambda$ for considering MRC for several values of $Nb_{lane}$. We can see an increases in the outage probability as the number of lanes increases. This results is intuitive because when the number of lanes increases the interfering vehicles density increases as well, hence increasing the outage probability.
Fig.\ref{Figure13(b)} shows NOMA outage probability as a function of $p$ using NOMA and considering the 1D-HPPP with a single lanes model, and the 1D-HPPP with multiple lanes. We can see from the Fig.\ref{Figure13(b)} that the single lane model matches perfectly the multiple lanes model.

\section{Conclusion}\label{Section7}

 In this paper, we implemented MRC using NOMA in VCs at road intersections.
We derived closed form expressions of the outage probability for a setup involving two destinations. Then we extended the analysis for a scenario involving $K$ destinations.
We also analyzed the performance for several road lanes.
We noticed that implementing MRC using NOMA in vehicles improvements significantly the performance.
compared to the standard cooperative transmission using NOMA.
We also noticed that MRC using NOMA significantly outperforms MRC using OMA.
From our results we concluded that it is always beneficial to use MRC and NOMA even at the cost of implementation complexity. 
Finally, we demonstrated that the outage probability has a peak when the vehicles are at the intersection, and that using MRC considering NOMA offers a great improvement in this context.

\appendix
\section{}\label{appA}

The outage probability related to $D_1$, denoted $\mathbb{P}(\textit{O}_{(1)})$, is expressed as
\begin{equation}
\mathbb{P}(\textit{O}_{(1)})=\mathbb{P}\Big( \mathcal{B}_{D_1} \cap \mathcal{A}_{D_1} \Big)+\mathbb{P}\Big(\mathcal{A}_{D_1}^C \cap \mathcal{C}_{D_1} \Big).
\end{equation}
First, we calculate the probability $\mathbb{P}\Big(\mathcal{A}_{R_1}^C \cap \mathcal{C}_{D_1} \Big)$ as follows
\begin{align}\label{chapter2c:equ1}
&\mathbb{P}\Big(\mathcal{A}_{R_1}^C \cap \mathcal{C}_{D_1} \Big)\nonumber\\
& =\mathbb{E}_{I_{X},I_Y}\Bigg[\mathbb{P}\Bigg\lbrace\ \frac{P\vert h_{SR}\vert^{2}l_{SR}a_1}{P\vert h_{SR}\vert^{2}l_{SR}a_2+I_{X_{R}}+I_{Y_{R}}+\sigma^2} \ge \Theta_1\nonumber\\ 
&\quad\quad\quad\bigcap \frac{P\left(\vert h_{SD_1}\vert^{2}l_{SD_1}+\vert h_{RD_1}\vert^{2}l_{RD_1}\right)\,a_1}{P\left(\vert h_{SD_1}\vert^{2}l_{SD_1}+\vert h_{RD_1}\vert^{2}l_{RD_1}\right)\,a_2+I_{X_{D_1}}+I_{Y_{D_1}}+\sigma^2} < \Theta_1\Bigg\rbrace\Bigg] \nonumber \\
=& \mathbb{E}_{I_{X},I_Y}\Bigg[\mathbb{P}\Bigg\lbrace\ P \vert h_{SR}\vert^{2}l_{SR}(a_1-\Theta_1 a_2)\ge\Theta_1\big[I_{X_{R}}+I_{Y_{R}}+\sigma^2\big]\nonumber\\ 
& \quad\bigcap P \left(\vert h_{SD_1}\vert^{2}l_{SD_1}+\vert h_{RD_1}\vert^{2}l_{RD_1}\right)\,(a_1-\Theta_1 a_2) < \Theta_1\big[I_{X_{D_1}}+I_{Y_{D_1}}+\sigma^2\big]\Bigg\rbrace\Bigg]. \nonumber\\
\end{align}

When $\Theta_1 < a_1/ a_2$, and after setting $G_{1}= \Theta_1 /(a_1- \Theta_1 a_2)$, then
\begin{flalign}\label{eq44}
&\mathbb{P}\Big(\mathcal{A}_{R_1}^C \cap \mathcal{C}_{D_1} \Big)\nonumber
\\ &=\mathbb{E}_{I_{X},I_Y}\Bigg[\mathbb{P}\Bigg\lbrace\ \vert h_{SR}\vert^{2}\ge \frac{G_1}{l_{SR}}\big[I_{X_{R}}+I_{Y_{R}}+\sigma^2/P\big] \nonumber\\ 
&\qquad \qquad  \bigcap \left(\vert h_{SD_1}\vert^{2}l_{SD_1}+\vert h_{RD_1}\vert^{2}l_{RD_1}\right) < G_1\big[I_{X_{D_1}}+I_{Y_{D_1}}+\sigma^2/P\big]\Bigg\rbrace\Bigg].&
\end{flalign}
Since $|h_{SR}|^2$ follows an exponential distribution with unit mean, we get
\begin{multline}\label{chapter2c:eq44}
\mathbb{P}\Big(\mathcal{A}_{R_1}^C \cap \mathcal{C}_{D_1} \Big)= \mathbb{E}_{I_{X},I_Y}\Bigg[\mathbb{P}\Bigg\lbrace\ \exp\left( \frac{G_1}{l_{SR}}\big[I_{X_{R}}+I_{Y_{R}}+\sigma^2/P\big]\right)\Bigg\rbrace\\ \times 1-\Bigg\lbrace \mathbb{P}\left(\vert h_{SD_1}\vert^{2}l_{SD_1}+\vert h_{RD_1}\vert^{2}l_{RD_1}\right) \ge G_1\big[I_{X_{D_1}}+I_{Y_{D_1}}+\sigma^2/P\big]\Bigg\rbrace\Bigg].
\end{multline}

We write the second probability in (\ref{chapter2c:eq44}) as
\begin{equation} \label{chapter2c:eq45} 
\mathbb{P}(\delta  \ge G_1 [I_{X_{D_1}}+I_{Y_{D_1}}+\sigma^2/P]),\nonumber
\end{equation}
where $\delta=\vert h_{RD_1}\vert^{2}l_{RD_1}+\vert h_{SD_1}\vert^{2}l_{SD_1}$. \\
The complementary cumulative distribution function of the random variable $\delta$, denoted $\bar{F}_\delta(.)$, is given by
\begin{center} \label{eq46} 
$\bar{F}_\delta(u)=\dfrac{l_{RD_1} e^{-u/l_{RD_1}}-l_{SD_1} e^{-u/l_{SD_1}}}{l_{RD_1}-l_{SD_1}}$.
\end{center}

Then, we have
\begin{multline} \label{chapter2c:eq47} 
\mathbb{P}\big[\vert h_{SD_1}\vert^{2}l_{SD_1}+\vert h_{RD_1}\vert^{2}l_{RD_1} \ge G_1 (I_{X_{D_1}}+I_{Y_{D_1}}+\sigma^2/P)\big]=\\
\dfrac{l_{RD_1} \exp\Big[- \dfrac{G_1}{l_{RD_1}}(I_{X_{D_1}}+I_{Y_{D_1}}+\sigma^2/P)\Big]- l_{SD_1}\exp\Big(- \dfrac{G_1}{l_{SD_1}}[I_{X_{D_1}}+I_{Y_{D_1}}+\sigma^2/P)\Big]}{l_{RD_1}-l_{SD_1}},
\end{multline}

Plugging (\ref{chapter2c:eq47}) into (\ref{chapter2c:eq44}) yields
\begin{align}
&\mathbb{P}\Big(\mathcal{A}_{R_1}^C \cap \mathcal{C}_{D_1} \Big)=\nonumber\\
&\mathbb{E}_{I_{X},I_Y}\Bigg[ \exp\left( \frac{G_1}{l_{SR}}\big[I_{X_{R}}+I_{Y_{R}}+\sigma^2/P\big]\right)\nonumber\\ 
&\qquad\times \Bigg\lbrace 1-\dfrac{l_{RD_1} \exp\Big(- \dfrac{G_1}{l_{RD_1}}(I_{X_{D_1}}+I_{Y_{D_1}}+\sigma^2/P)\Big)}{l_{RD_1}-l_{SD_1}} \nonumber 
\\ &\qquad - \dfrac{l_{SD_1}\exp\Big(- \dfrac{G_1}{l_{SD_1}}[I_{X_{D_1}}+I_{Y_{D_1}}+\sigma^2/P]\Big)}{l_{RD_1}-l_{SD_1}}\Bigg\rbrace\Bigg]
\nonumber\\
=& \mathbb{E}_{I_{X},I_Y}\Bigg[ \exp\left( \frac{G_1}{l_{SR}}\big[I_{X_{R}}+I_{Y_{R}}+\sigma^2/P\big]\right)\nonumber -  \exp\left( \frac{G_1}{l_{SR}}\big[I_{X_{R}}+I_{Y_{R}}+\sigma^2/P\big]\right) \nonumber\\ 
&\qquad \times \dfrac{l_{RD_1} \exp\Big(- \dfrac{G_1}{l_{RD_1}}[I_{X_{D_1}}+I_{Y_{D_1}}+\sigma^2/P]\Big)}{l_{RD_1}-l_{SD_1}}
\nonumber\\
&\qquad - \dfrac{ l_{SD_1}\exp\Big(- \dfrac{G_1}{l_{SD_1}}[I_{X_{D_1}}+I_{Y_{D_1}}+\sigma^2/P]\Big)}{l_{RD_1}-l_{SD_1}}\big)\Bigg]\nonumber.
\end{align}

Given that  $\mathbb{E}[e^{sI}]=\mathcal{L}_I(s)$,  we finally get
\begin{flalign}
\mathbb{P}\Big(\mathcal{A}_{R_1}^C \cap \mathcal{C}_{D_1} \Big)=&
\mathcal{L}_{I_{X_{R}}}\bigg(\frac{G_{1}}{l_{SR}}\bigg)\mathcal{L}_{I_{Y_{R}}}\bigg(\frac{G_{1}}{l_{SR}}\bigg)\exp\bigg(-\frac{\sigma^2 G_{1}}{P l_{SR}}\bigg)\nonumber\\
&-\mathcal{L}_{I_{X_{R}}}\bigg(\frac{G_{1}}{l_{SR}}\bigg)\mathcal{L}_{I_{Y_{R}}}\bigg(\frac{G_{1}}{l_{SR}}\bigg)\exp\bigg(-\frac{\sigma^2 G_{1}}{P l_{SR}}\bigg)\nonumber\\
&\frac{l_{RD_1}\mathcal{L}_{I_{X_{D_1}}}\bigg(\frac{G_{1}}{l_{RD_1}}\bigg)\mathcal{L}_{I_{Y_{D_1}}}\bigg(\frac{G_{1}}{l_{RD_1}}\bigg)\exp\bigg(-\frac{\sigma^2 G_{1}}{P l_{RD_1}}\bigg)}{l_{RD_1}-l_{SD_1}}\nonumber\\
&-\frac{l_{SD_1}\mathcal{L}_{I_{X_{D_1}}}\bigg(\frac{G_{1}}{l_{SD_1}}\bigg)\mathcal{L}_{I_{Y_{D_1}}}\bigg(\frac{G_{1}}{l_{SD_1}}\bigg)\exp\bigg(-\frac{\sigma^2 G_{1}}{P l_{SD_1}}   \bigg)}{l_{RD_1}-l_{SD_1}} .
\end{flalign}

The probability $\mathbb{P}\Big( \mathcal{B}_{D_1} \cap \mathcal{A}_{D_1} \Big)$ can be expressed as 
\begin{eqnarray}\label{chapter2c:quat1}
\mathbb{P}\Big( \mathcal{B}_{D_1} \cap \mathcal{A}_{D_1} \Big)&=&1-\mathbb{P}\Big( \mathcal{B}_{D_1}^C \cup \mathcal{A}_{D_1}^C \Big)\nonumber\\
&=&1-\mathbb{P}\Big(\mathcal{B}_{D_1}^C\Big)-\mathbb{P}\Big(\mathcal{A}_{D_1}^C\Big)+\mathbb{P}\Big(\mathcal{B}_{D_1}^C \cap \mathcal{A}_{D_1}^C\Big).
\end{eqnarray}
The probabilities in (\ref{chapter2c:quat1}) can be calculated following the same steps above.

In the same way, we calculate $\mathbb{P}(\textit{O}_{(2)})$ as 
\begin{eqnarray}\label{chapter2c:quat2}
\mathbb{P}(\textit{O}_{(2)})&=& \mathbb{P}\left[\Bigg\{\bigcup_{i=1}^{2} \mathcal{B}_{D_{2-i}}(\Theta_i)\Bigg\}\cap\Bigg\{\ \bigcup_{i=1}^{2} \mathcal{A}_{R_i}(\Theta_i)\Bigg\}\right] 
\nonumber\\&& + \mathbb{P}\left[\Bigg\{\bigcap_{i=1}^{2} \mathcal{A}_{R_{i}}^C(\Theta_i)\Bigg\} \cap \Bigg\{ \bigcup_{i=1}^{2} \mathcal{C}_{D_{2-i}}(\Theta_i)\Bigg\}\right].  
\end{eqnarray}
To calculate the first probability in (\ref{chapter2c:quat2}), we proceed as follows
\begin{align}\label{chapter2c:quat3}
&\mathbb{P}\left[\Bigg\{\bigcup_{i=1}^{2} \mathcal{B}_{D_{2-i}}(\Theta_i)\Bigg\}\cap \Bigg\{\bigcup_{i=1}^{2} \mathcal{A}_{R_i}(\Theta_i)\Bigg\}\right]\nonumber\\
&=1- \mathbb{P}\left[\Bigg\{\bigcap_{i=1}^{2} \mathcal{B}_{D_{2-i}}^C(\Theta_i)\Bigg\}\cup \Bigg\{\bigcap_{i=1}^{2} \mathcal{A}_{R_i}^C(\Theta_i)\Bigg\}\right]\nonumber\\
&=1- \mathbb{P}\left[\bigcap_{i=1}^{2} \mathcal{B}_{D_{2-i}}^C(\Theta_i)\right]- \mathbb{P}\left[\bigcap_{i=1}^{2} \mathcal{A}_{R_{i}}^C(\Theta_i)\right]\nonumber\\
&+\mathbb{P}\left[\Bigg\{\bigcap_{i=1}^{2} \mathcal{B}_{D_{2-i}}^C(\Theta_i)\Bigg\}\cap \Bigg\{\bigcap_{i=1}^{2} \mathcal{A}_{R_i}^C(\Theta_i)\Bigg\}\right].\nonumber\\
\end{align}
The first two probabilities in (\ref{chapter2c:quat3}) can be calculated in a straightforward manner as above. The last probability in (\ref{chapter2c:quat3}), that we denote by $\mathcal{P}_1$, is expressed as 
\begin{align}
\mathcal{P}_1 =&
\mathbb{E}_{I_{X},I_Y}\Bigg[\mathbb{P}\Bigg\lbrace\frac{P\vert h_{SD_2}\vert^{2}l_{SD_2}a_1}{P\vert h_{SD_2}\vert^{2}l_{SD_2}a_2+I_{X_{D_2}}+I_{Y_{D_2}}+\sigma^2} \ge \Theta_1\nonumber\\
&\qquad \qquad \bigcap 
\frac{P\vert h_{SD_2}\vert^{2}l_{SD_2}a_2}{\beta P\vert h_{SD_2}\vert^{2}l_{SD_2}a_1+I_{X_{D_2}}+I_{Y_{D_2}}+\sigma^2} \ge \Theta_2\nonumber\\
& \qquad \qquad\bigcap \frac{P\vert h_{SR}\vert^{2}l_{SR}a_1}{P\vert h_{SR}\vert^{2}l_{SR}a_2+I_{X_{R}}+I_{Y_{R}}+\sigma^2} \ge \Theta_1\nonumber\\
& \qquad \qquad  \bigcap
\frac{P\vert h_{SR}\vert^{2}l_{SR}a_2}{\beta P\vert h_{SR}\vert^{2}l_{SR}a_1+I_{X_{R}}+I_{Y_{R}}+\sigma^2} \ge \Theta_2\Bigg\rbrace\Bigg].\nonumber
\end{align}
When $\Theta_1 < a_1/ a_2$ and $\Theta_2 < a_2/ \beta a_1$, we get
\begin{align}
\mathcal{P}_1 =&
\mathbb{E}_{I_{X},I_Y}\Bigg[\mathbb{P}\Bigg\lbrace\vert h_{SD_2}\vert^{2}\ge \frac{G_1}{l_{SD_2}}\big[I_{X_{D_2}}+I_{Y_{D_2}}+\sigma^2/P\big]\nonumber\\
&\qquad \qquad\bigcap 
\vert h_{SD_2}\vert^{2}\ge \frac{G_2}{l_{SD_2}}\big[I_{X_{D_2}}+I_{Y_{D_2}}+\sigma^2/P\big]\nonumber\\
&\qquad \qquad\bigcap \vert h_{SR}\vert^{2}\ge \frac{G_1}{l_{SR}}\big[I_{X_{R}}+I_{Y_{R}}+\sigma^2/P\big]
\nonumber\\&\qquad \qquad\bigcap 
\vert h_{SR}\vert^{2}\ge \frac{G_2}{l_{SR}}\big[I_{X_{R}}+I_{Y_{R}}+\sigma^2/P\big]\Bigg\rbrace\Bigg].\nonumber\\
=&\mathbb{E}_{I_{X},I_Y}\Bigg[\mathbb{P}\Bigg\lbrace\vert h_{SD_2}\vert^{2}\ge \frac{\max(G_1,G_2)}{l_{SD_2}}\big[I_{X_{D_2}}+I_{Y_{D_2}}+\sigma^2/P\big] \nonumber\\
&\qquad \qquad\bigcap \vert h_{SR}\vert^{2}\ge \frac{\max(G_1,G_2)}{l_{SR}}\big[I_{X_{R}}+I_{Y_{R}}+\sigma^2/P\big]\Bigg\rbrace\Bigg].
\end{align}
In the case when $G_2=\Theta_2 /(a_2-\Theta_2 \beta a_1)$,

Finally, we obtain
\begin{align}
&\mathbb{P}\left[\Bigg\{\bigcap_{i=1}^{2} \mathcal{B}_{D_{2-i}}^C(\Theta_i)\Bigg\}\cap \Bigg\{\bigcap_{i=1}^{2} \mathcal{A}_{R_i}^C(\Theta_i)\Bigg\}\right]=\nonumber\\
&\mathcal{L}_{I_{X_{D_2}}}\bigg(\frac{G_{\mathrm{max}}}{l_{SD_2}}\bigg)\mathcal{L}_{I_{Y_{D_2}}}\bigg(\frac{G_{\mathrm{max}}}{l_{SD_2}}\bigg)\exp\bigg(-\frac{\sigma^2 G_{\mathrm{max}}}{P l_{SD_2}}   \bigg)\nonumber\\
&\times\mathcal{L}_{I_{X_{R}}}\bigg(\frac{G_{\mathrm{max}}}{l_{SR}}\bigg)\mathcal{L}_{I_{Y_{R}}}\bigg(\frac{G_{\mathrm{max}}}{l_{SR}}\bigg)\exp\bigg(-\frac{\sigma^2 G_{\mathrm{max}}}{P l_{SR}}   \bigg),\nonumber\\
\end{align}
where $G_{\mathrm{max}}=\mathrm{max}(G_1,G_2)$.\\
The second probability in (\ref{chapter2c:quat2}) can be calculated following the same steps above.

\section{}\label{chapter3:App4} 
The expression of the Laplace transform of interference originated from the $X_2$ road at $M$ is given by
\begin{equation}\label{aq1}
\mathcal{L}_{I_{{X_2}_M}}(s)=\exp\Bigg(-\emph{p}\lambda_{X_2}\int_{-\infty}^{+\infty}\dfrac{1}{1+\dfrac{(\Vert x-M \Vert^\alpha)}{s}}dx\Bigg)
\end{equation}
where
\begin{equation}\label{}
\Vert x-M \Vert=\sqrt{m_{y_2}^2+(x-m_{x_2})^2 }
\end{equation}
and $m_{x_2}$ and $m_{y_2}$ are the coordinate of $M$ at the $X_2$ and $Y_2$ road.

For $\alpha=2$, (\ref{aq1}) becomes
\begin{equation}\label{aq2}
\mathcal{L}_{I_{{X_2}_M}}(s)=\exp\Bigg(-\emph{p}\lambda_{X_2}s\int_{-\infty}^{+\infty}\dfrac{1}{s+m_{y_2}^2+(x-m_{x_2})^2}dx\Bigg)
\end{equation}
and the integral inside the exponential in (\ref{aq2}) equals:
\begin{equation}\label{aq5}
\int_{-\infty}^{+\infty}\dfrac{1}{s+m_{y_2}^2+(x-m_{x_2})^2}dx=\dfrac{\pi}{s+m_{y_2}^2}
\end{equation}
We express $m_{x_2}$ and $ m_{y_2}$ as a function of $m$ and $\theta_M $ as follows
\begin{equation}\label{aq3}
m_{x_2}=m \cos(\theta_M)-d_{X_{Road}}
\end{equation}
and 
\begin{equation}\label{aq4}
 m_{y_2}= m \sin(\theta_M)-d_{Y_{Road}}
\end{equation}

Substituting (\ref{aq4}) in (\ref{aq5}), then in (\ref{aq2}) yields (\ref{49a}). Following the same steps we obtain (\ref{50a}).

%
%
%
%
%


\section*{\refname}
\bibliography{bibnoma}
\end{document}